\documentclass[preprint,10pt,sort&compress]{elsarticle}
\usepackage{amssymb}
\usepackage{amsmath}
\usepackage{amsthm}
\usepackage{amsfonts}
\usepackage{amssymb}
\usepackage{graphicx}

\journal{Physica A}

\begin{document}

\begin{frontmatter}

\title{Quasi-stationary Chaotic States in Multi-dimensional Hamiltonian Systems}

\author[TEI,UoP]{Ch. Antonopoulos\corref{cor1}}
\ead{chantonopoulos@teimes.gr}

\author[UoP]{T. Bountis}
\ead{bountis@math.upatras.gr}

\author[ULB]{V. Basios}
\ead{vbasios@ulb.ac.be}

\address[TEI]{Department of Automation and\\High Performance Computing Systems - Programming \& Algorithms Lab (HPCS Lab), Technological Educational Institute of Messolonghi, Nea Ktiria, 30200\\Messolonghi, Greece}

\address[UoP]{Center for Research and Applications of Nonlinear Systems (CRANS), Department of Mathematics, University of Patras, 265$\,$00, Patras, Greece}

\address[ULB]{Interdisciplinary Center for Nonlinear Phenomena and Complex Systems (CENOLI), Service de Physique des Syst\`{e}mes Complexes et M\'{e}canique Statistique, Universit\'{e} Libre de Bruxelles, 1050, Brussels, Belgium}

\cortext[cor1]{Corresponding author}

%\biboptions{longnamesfirst,angle,semicolon}

\begin{abstract}
We study numerically statistical distributions of sums of chaotic orbit coordinates, viewed as independent random variables, in weakly chaotic regimes of three multi-dimensional Hamiltonian systems: Two Fermi-Pasta-Ulam (FPU-$\beta$) oscillator chains with different boundary conditions and numbers of particles and a microplasma of identical ions confined in a Penning trap and repelled by mutual Coulomb interactions. For the FPU systems we show that, when chaos is limited within ``small size'' phase space regions, statistical distributions of sums of chaotic variables are well approximated for surprisingly long times (typically up to $t\approx10^6$) by a $q$-Gaussian ($1<q<3$) distribution and tend to a Gaussian ($q=1$) for longer times, as the orbits eventually enter into ``large size'' chaotic domains. However, in agreement with other studies, we find in certain cases that the $q$-Gaussian is not the only possible distribution that can fit the data, as our sums may be better approximated by a different so-called ``crossover'' function attributed to finite-size effects. In the case of the microplasma Hamiltonian, we make use of these $q$-Gaussian distributions to identify two energy regimes of ``weak chaos''-one where the system melts and one where it transforms from liquid to a gas state-by observing where the $q$-index of the distribution increases significantly above the $q=1$ value of strong chaos.
\end{abstract}


\end{frontmatter}

%\pacs{05.45.-a,05.45.Jn,05.45.Pq}

%%%%%%%%%%%%%%%%%%%%%%%%%%%%%%%%%%%%%%%%%%%%%%%%%%%%%%%%%%%%%%%%%%%%%%%%%%%%%%%%%%%%%%%%%%%%%%%%%%%%%%%%%%%%%%%%%%%%%%%%%%%%%%%%%%
%%%%%%%%%%%%%%%%%%%%%%%%%%%%%%%%%%%%%%%%%%%%%%%%%%%%%%%%%%%%%%%%%%%%%%%%%%%%%%%%%%%%%%%%%%%%%%%%%%%%%%%%%%%%%%%%%%%%%%%%%%%%%%%%%%
%%%%%%%%%%%%%%%%%%%%%%%%%%%%%%%%%%%%%%%%%%%%%%%%%%%%%%%%%%%%%%%%%%%%%%%%%%%%%%%%%%%%%%%%%%%%%%%%%%%%%%%%%%%%%%%%%%%%%%%%%%%%%%%%%%

\section{Introduction}\label{intro}

Probability density functions (pdfs) of chaotic trajectories of dynamical systems have been studied for many decades and by many authors, aiming to understand the transition from deterministic to stochastic dynamics \cite{Anosov1967,Arnold1967,Sinai1972,Pesin1976,Pesin1977,Ruelle1979,katok1980,Ruelle1980,Ruelle1982,Eckmann1985,katok1985}. The fundamental question in this context concerns the existence of an appropriate invariant probability density (or ergodic measure), characterizing chaotic motion in phase space regions where solutions generically exhibit exponential divergence of nearby trajectories. If such an invariant measure can be established for almost all initial conditions (i.e. except for a set of measure zero), one has a firm basis for studying the system from a Statistical Mechanics point of view.

If, additionally, this invariant measure turns out to be a continuous and sufficiently smooth function of the phase space coordinates, one can invoke the Boltzmann-Gibbs (BG) microcanonical ensemble and attempt to evaluate all relevant quantities of equilibrium Statistical Mechanics, like partition function, free energy, entropy, etc. of the system. On the other hand, if the measure is absolutely continuous (as e.g. in the case of the so-called Axiom A dynamical systems), one might still be able to use the formalism of modern ergodic theory and SRB measures to study the statistical properties of the model \cite{Eckmann1985}.

In all these cases, viewing the values of one (or a linear combination) of components of a chaotic solution at discrete times $t_n, n=1,\ldots,\mathcal{N}$ as realizations of $\mathcal{N}$ independent and identically distributed random variables $X_n$ and calculating the distribution of their sums, one expects to find a Gaussian distribution, whose mean and variance are those of the $X_n$'s. This is indeed what happens for many chaotic dynamical systems studied to date which are \textit{ergodic}, i.e. almost all their orbits (except for a set of measure zero) pass arbitrarily close to any point of the constant energy manifold, after sufficiently long enough times. What is also true is that, in these cases, at least one Lyapunov exponent is positive, stable periodic orbits are absent and the constant energy manifold is covered uniformly by chaotic orbits, for all but a (Lebesgue) measure zero set of initial conditions.

But then, what about ``small size'' chaotic regions of Hamiltonian systems, at energies where the maximal Lyapunov exponent is positive (but still rather ``small'') and stable periodic orbits still exist, whose islands of invariant tori and sets of cantori surrounding them occupy a positive measure subset of the energy manifold? In such regimes of so-called ``weak chaos'', a great number of orbits stick for long times to the boundaries of these islands and chaotic trajectories diffuse slowly through multiply connected regions in a highly non-uniform way  \cite{Aizawa1984,Chirikov1984,Meiss1986}. Such examples occur in many physically realistic systems studied in the current literature (see e.g. \cite{Skokos2008,Flach2009,Johansson2009,Skokos2009}).

In this paper, we focus on such ``weakly chaotic'' regimes and demonstrate by means of numerical experiments in the spirit of the Central Limit Theorem \cite{Rice1995} that pdfs of sums of orbit components \textit{do not} rapidly converge to a Gaussian, but are well approximated, for long integration times, by the so-called $q$-Gaussian distribution \cite{Tsallisbook2009}
\begin{equation}\label{q_gaussian}
P(s)=a \exp_q({-\beta s^2})\equiv a\biggl[1-(1-q)\beta s^{2}\biggr]^{\frac{1}{1-q}}
\end{equation}
where the index satisfies $1<q<3$, $\beta$ is an arbitrary parameter and $a$ is a normalization constant. At longer times, of course, chaotic orbits eventually seep out from smaller regions to larger chaotic seas, where obstruction by islands and cantori is less dominant and the dynamics is more uniformly ergodic. This transition is signaled by the $q$-index of the distribution (\ref{q_gaussian}) decreasing towards $q=1$, which represents the limit at which it becomes a Gaussian.

Thus, in our models, $q$-Gaussian distributions represent \textit{quasi-stationary states} (QSS) that are often very \textit{long-lived}, especially inside ``thin'' chaotic layers near periodic orbits that have \textit{just turned unstable}. This suggests that it might be useful to study these pdfs (as well as their associated $q$ values) for sufficiently long times and try to derive useful information about the corresponding QSS directly from the chaotic orbits of the system, at least for time intervals accessible by numerical integration. QSS have already been studied in coupled standard maps and the so-called HMF model in \cite{Baldovin2004a,Baldovin2004b}, but not from the viewpoint of sum distributions. These authors calculated the growth of the variance of the total ``angular momentum'' of their models and obtained interesting results, which we shall discuss in the Conclusions section of the present paper.

Care must be taken, however, with regard to the kind of functions that are used to approximate these pdfs. While it is true that $q$-Gaussians offer a popular and quite accurate representation of QSS sum distributions, they are \textit{not} the only possible choice. As we have found in certain cases, other so-called \textit{``crossover'' functions} \cite{Tsallisbook2009,CelikogluTirnakli2010,TsallisTirnakli2010} (see e.g. Eq. (\ref{eq28paperTsallisTirnakli}) below) can be used to describe the same data with better accuracy. Thus, as pointed out already by several authors \cite{Dauxois2007,HilhorstSchehr2007,Hilhorst2010}, there may be \textit{other alternatives} to $q$-Gaussians, that can approximate better sum distributions of weakly chaotic QSS such as the ones studied here.

In what follows, after some preliminaries about our method in Section \ref{CLT_approach}, we begin in Section \ref{FPU_OPM_PBC_section} by studying the solutions of a Fermi-Pasta-Ulam (FPU)-$\beta$ 1-dimensional chain of $N=128$ particles, under \textit{periodic} boundary conditions, near one of its simple periodic orbits called the $\pi$-mode, at energies where it has just turned unstable. This is one of the \textit{nonlinear normal modes} (NNMs) of the chain and represents a continuation of the linear mode that consists of all neighboring pairs of particles oscillating with the same amplitude and a $\pi$-phase difference with respect to each other \cite{Budinsky1983,Poggi1997,Dauxois1997,Cafarella2004,Antonopoulos2006IJBC,Bountis2010}.

Following the approach of \cite{Leo2010}, we verify that the pdfs of chaotic orbits starting very close to the unstable $\pi$-mode are well approximated by a $q$-Gaussian over time intervals of the order of $t\approx 10^6$. These trajectories, however, represent a QSS: Their distributions for longer times deviate from a $q$-Gaussian and show a tendency to converge to a Gaussian as $q\rightarrow 1$. Remarkably enough, this result holds true both for the distribution of the \textit{single values} of the chaotic variable (as shown recently in \cite{Leo2010}), as well as the distribution of its \textit{sums} relevant to our study.

In seeking to relate our statistical results to the actual dynamics of the chain, we adopt a very interesting approach described by Cretegny et al. in \cite{Cretegnyetal1998} and compute their $C_0(t)$ function for initial conditions close to the $\pi$-mode at low energies, just above the destabilization threshold of the mode (always under periodic boundary conditions). We find, just as in \cite{Cretegnyetal1998}, that $C_0(t)$ displays a maximum as a chaotic breather appears on the chain, until $t\approx10^8$, when the breather collapses and energy equipartition occurs. Remarkably, this is about the time when our $q$-Gaussian approximations converge to Gaussians. On the other hand, for an FPU chain with $N=129$ particles under \textit{fixed} boundary conditions the $C_0(t)$ function does \textit{not} exhibit a clear maximum and no chaotic breather is observed, while a $q$-Gaussian approximation is still valid for long times, eventually turning into a Gaussian when $C_0(t)$ settles down to its limiting value and energy is equally distributed among all degrees of freedom.

We then turn, in Section \ref{FPU_SPO1_FBC_section}, to a similar investigation regarding an FPU-$\beta$ model with only $N=5$ particles and \textit{fixed} boundary conditions. We first choose a NNM called the SPO1 mode where, between every two particles oscillating with the same amplitude and opposite phase, there is one that is stationary for all $t$. Starting again with a total energy just above the (first) destabilization threshold and choosing initial conditions in the vicinity of the NNM, we discover that the sum distributions of chaotic orbits have again a $q$-Gaussian shape, with $q$ significantly higher than 1. Integrating the equations of motion for times of the order of $t=10^6$, we observe that the orbits form a thin ``figure eight'' in a 2-dimensional $(q_1,p_1)$ projection of the energy manifold. In fact, the closer one starts to the NNM, the longer the persistence of the $q$-Gaussian QSS and the thinner the width of the ``figure eight'' region.

However, as soon as we start deviating significantly from the unstable periodic point, the transitory nature of the QSS becomes apparent: Already for time intervals of the order of $t=10^5$ or $t=10^6$, sum distributions begin to show a clear tendency to approach a Gaussian. Furthermore, the ``figure eight''-like region expands to larger domains of phase space, where the Lyapunov exponents increase by orders of magnitude and the motion is more ``uniformly'' chaotic. Analogous results in this direction have also been obtained very recently on low-dimensional conservative maps in \cite{RuizBountisTsallis2010}.

We then carry out a similar study near another NNM (i.e. the so-called SPO2 mode), where every two oppositely moving particles one is stationary for all times. In this case we find QSS that are much longer-lived, possessing sum pdfs of the $q$-Gaussian type that do not tend to a Gaussian even for times as long as $t=10^{10}$! Here, chaotic orbits trace out thin regions of nearly vanishing Lyapunov exponents close to banana-shaped tori representing a true ``edge of chaos'' regime. The interesting observation here is that the pdf we obtain in the long time limit converges to a \textit{smooth function} that is well approximated by the ``crossover'' formula (\ref{eq28paperTsallisTirnakli}) proposed to hold in a regime between $q$-Gaussians and Gaussians, where finite size (and time) effects yield distributions with lower tails \cite{Tsallisbook2009,CelikogluTirnakli2010,TsallisTirnakli2010}.

Finally, we turn in Section \ref{microplasma_section} to a very different Hamiltonian system that describes $N$ identical ions in a microplasma, mutually repelled by Coulomb interactions and confined radially by a magnetic field and axially by an electrostatic field (Penning trap). Here the interactions are \textit{long range}, unlike the FPU models where only nearest neighbors are involved. In a recent publication \cite{Antonopoulos2010PRE}, we have shown that there is a critical energy range, $2\leq E\leq 4$, where $N=5$ ions undergo a ``melting transition'' \cite{Hill1994}, as they exit the domain of their local potential minima and begin to wander chaotically throughout the available phase space. Interestingly enough, when one uses $q$-Gaussian distributions to approximate the statistics of orbits in this region, one discovers that, for $2<E<3$, the $q$ values steadily increase above 1 and attain a peak value of $q\approx 1.8$ (near $E\approx 2.8$) before returning to $q=1$ at higher energies of much larger positive Lyapunov exponents and uniformly spread chaos. Performing then a similar analysis for $E>3$, we identify a \textit{second} much broader energy range (i.e. $E\in(30,200)$) over which the system undergoes a further transition, passing from a liquid to a gas state, where $q$ values attain another peak well above unity after which they start gradually to decrease and then fluctuate around the $q=1$ value of uniform chaos.

Regarding the computation of the Lyapunov exponents, we have applied throughout this work the well-known method of Benettin et. al \cite{Benettin1980a,Benettin1980b}, together with a strategy proposed in \cite{Skokosetal2010} to follow the time evolution of the deviation vectors. Finally, we conclude with a summary and discussion of our results in Section \ref{conclusions}.

%%%%%%%%%%%%%%%%%%%%%%%%%%%%%%%%%%%%%%%%%%%%%%%%%%%%%%%%%%%%%%%%%%%%%%%%%%%%%%%%%%%%%%%%%%%%%%%%%%%%%%%%%%%%%%%%%%%%%%%%%%%%%%%%%%
%%%%%%%%%%%%%%%%%%%%%%%%%%%%%%%%%%%%%%%%%%%%%%%%%%%%%%%%%%%%%%%%%%%%%%%%%%%%%%%%%%%%%%%%%%%%%%%%%%%%%%%%%%%%%%%%%%%%%%%%%%%%%%%%%%
%%%%%%%%%%%%%%%%%%%%%%%%%%%%%%%%%%%%%%%%%%%%%%%%%%%%%%%%%%%%%%%%%%%%%%%%%%%%%%%%%%%%%%%%%%%%%%%%%%%%%%%%%%%%%%%%%%%%%%%%%%%%%%%%%%

\section{Statistical distributions of chaotic QSS and their computation}\label{CLT_approach}

The problems we investigate in this paper are described by an autonomous $N$ degree of freedom Hamiltonian function of the form
\begin{equation}
H\equiv H(\mathbf{q}(t),\mathbf{p}(t))=H(q_1(t),\ldots,q_N(t),p_1(t),\ldots,p_N(t))=E\label{Ham_fun}
\end{equation}
where $(q_k(t), p_k(t))$ are the positions and momenta respectively representing the solution in phase space at time $t$. As is well-known, these solutions can be periodic, quasi-periodic or chaotic depending on the initial condition $(\mathbf{q}(0)$,$\mathbf{p}(0))$ and the value of the total energy $E$. What we wish to study here are the statistical properties of such systems in regimes of ``weakly'' chaotic motion, where the Lyapunov exponents \cite{Benettin1980a,Benettin1980b,Eckmann1985,Skokos2010} are positive but very small. Such situations often arise when one considers solutions which slowly diffuse into thin chaotic layers and wander through a complicated network of higher order resonances, often sticking for very long times to the boundaries of islands constituting the so-called ``edge of chaos'' regime \cite{Tsallisbook2009}.

There are several interesting questions to ask here: How long do these ``weakly'' chaotic states last? Assuming they are quasi-stationary, can we describe them \textit{statistically} by simply observing some of their chaotic orbits? What type of distributions characterize these QSS and how could one connect them to the actual dynamics of the solutions of the corresponding Hamiltonian system? Does the dimensionality of the Hamiltonian,
i.e. the number $N$ of degrees of freedom, play a role in these considerations?

The approach we shall follow is in the spirit of the well-known Central Limit Theorem \cite{Rice1995}. In particular, we shall use the solutions of Hamilton's equations of motion
\begin{equation}\label{eq:Hamiltonian_ODEs}
\frac{{dq}_{k}}{dt}=\frac{\partial H}{{\partial
p}_{k}},\qquad\frac{{dp}_{k}}{dt}=-\frac{\partial H}{{\partial
q}_{k}},\;k=1,\ldots,N
\end{equation}
to construct \textit{distributions} of suitably rescaled sums of $M$ values of a generic observable $\eta_i=\eta(t_i)\;(i=1,\ldots,M)$ which depends linearly on the components of the solution. If these are viewed as independent and identically distributed random variables (in the limit $M\rightarrow\infty$), we may evaluate their sum
\begin{equation}\label{sums_CLT}
S_M^{(j)}=\sum_{i=1}^M\eta_i^{(j)}
\end{equation}
for $j=1,\ldots,N_{\mbox{ic}}$ different initial conditions. Thus, we can study the statistics of the variables $S_M^{(j)}$, centered about their mean value $\langle S_M^{(j)}\rangle=\frac{1}{N_{\mbox{ic}}}\sum_{j=1}^{N_{\mbox{ic}}}\sum_{i=1}^{M}\eta_i^{(j)}$ and rescaled by their standard deviation $\sigma_M$
\begin{equation}
s_M^{(j)}\equiv\frac{1}{\sigma_M}\Bigl(S_M^{(j)}-\langle S_M^{(j)}\rangle \Bigr)=\frac{1}{\sigma_M}\Biggl(\sum_{i=1}^M\eta_i^{(j)}-\frac{1}{N_{\mbox{ic}}}\sum_{j=1}^{N_{\mbox{ic}}}\sum_{i=1}^{M}\eta_i^{(j)}\Biggl)
\end{equation}
where
\begin{equation}
\sigma_M^2=\frac{1}{N_{\mbox{ic}}}\sum_{j=1}^{N_{\mbox{ic}}}\Bigl(S_M^{(j)}-\langle S_M^{(j)}\rangle \Bigr)^2=\langle S_M^{(j)2}\rangle -\langle S_M^{(j)}\rangle^2.
\end{equation}
Plotting the normalized histogram of the probabilities $P(s_M^{(j)})$ as a function of $s_M^{(j)}$, we then compare our pdfs with a $q$-Gaussian of the form
\begin{equation}\label{q_gaussian_distrib}
P(s_M^{(j)})=a\exp_q({-\beta s_M^{(j)2}})\equiv a\biggl[1-(1-q)\beta s_M^{(j)2}\biggr]^{\frac{1}{1-q}}
\end{equation}
where $q$ is the so-called entropic index, $\beta$ is a free parameter and $a$ a normalization constant \cite{Tsallisbook2009}. Function (\ref{q_gaussian_distrib}) is a generalization of the well-known Gaussian distribution, since in the limit $q\rightarrow 1$ we have
$\exp_q(-\beta x^2)\rightarrow\exp(-\beta x^2)$.

Moreover, it can be shown that the $q$-Gaussian distribution (\ref{q_gaussian_distrib}) is normalized when
\begin{equation}\label{beta-$q$-Gaussian}
\beta=a\sqrt{\pi}\frac{\Gamma\Bigl(\frac{3-q}{2(q-1)}\Bigr)}{(q-1)^{\frac{1}{2}}\Gamma\Bigl(\frac{1}{q-1}\Bigr)}
\end{equation}
where $\Gamma$ is the Euler $\Gamma$ function. Clearly, Eq. (\ref{beta-$q$-Gaussian}) shows that the allowed values of $q$ are $1<q<3$.

The index $q$ appearing in Eq. (\ref{q_gaussian_distrib}) is connected with the Tsallis entropy \cite{Tsallisbook2009}
\begin{equation}\label{Tsallis entropy}
S_q=k\frac{1-\sum_{i=1}^W p_i^q}{q-1}\mbox{ with }\sum_{i=1}^W p_i=1
\end{equation}
where $i=1,\ldots,W$ counts the microstates of the system, each occurring with a probability $p_i$ and $k$ is the so-called Boltzmann universal constant. Just as the Gaussian distribution represents an extremal of the BG entropy $S_{\mbox{BG}}\equiv S_1=k\sum_{i=1}^W p_i\ln p_i$, so is the $q$-Gaussian (\ref{q_gaussian_distrib}) derived by optimizing the Tsallis entropy (\ref{Tsallis entropy}) under appropriate constraints.

Systems characterized by the Tsallis entropy are said to lie at the ``edge of chaos'' and are significantly different than BG systems, in the sense that their entropy is nonadditive and generally nonextensive \cite{Tsallisbook2009,TsallisTirnakli2010}. In fact, a $q$-Central Limit Theorem was proved \cite{UmarovTsallis2008} for $q$-Gaussian distributions (\ref{q_gaussian_distrib}). However, the role of this theorem in establishing the theoretical foundation of such functions as limiting distributions of systems of strongly correlated variables has been recently challenged in \cite{Hilhorst2010}.

Let us now describe the numerical approach we are going to follow to calculate the above pdfs. First, we specify an observable denoted by $\eta(t)$ as one (or a linear combination) of the components of the position vector $\mathbf{q}(t)$ of a chaotic solution of Eq. (\ref{eq:Hamiltonian_ODEs}), located initially at $(\mathbf{q}(0),\mathbf{p}(0))$.

Assuming that the orbit visits all parts of a QSS during the integration interval $0\leq t\leq t_{\mbox{f}}$, we divide it into $N_{\mbox{ic}}$ equally spaced, consecutive time windows, which are long enough to contain a significant part of the orbit. Next, we subdivide each such window into a number $M$ of equally spaced subintervals and calculate the sum $S_M^{(j)}$ of the values of the observable $\eta(t)$ at the \textit{right edges} of these subintervals (see Eq. (\ref{sums_CLT})).

In this way, we treat the point at the beginning of every time window as a new initial condition and repeat this process $N_{\mbox{ic}}$ times to obtain as many sums as required for reliable statistics. Consequently, at the end of the integration, we compute the average and standard deviation of the sums (\ref{sums_CLT}), evaluate the $N_{\mbox{ic}}$ rescaled quantities $s_M^{(j)}$ and plot the histogram $P(s_M^{(j)})$ of their distribution.

As we shall see in the next sections, in regions of ``weak chaos'' these distributions are well-fitted by a $q$-Gaussian of the form (\ref{q_gaussian_distrib}) for fairly long time intervals. However, for longer times, as the orbits begin to diffuse to wider domains of ``strong chaos'', the well-known form of a Gaussian pdf is recovered.

%%%%%%%%%%%%%%%%%%%%%%%%%%%%%%%%%%%%%%%%%%%%%%%%%%%%%%%%%%%%%%%%%%%%%%%%%%%%%%%%%%%%%%%%%%%%%%%%%%%%%%%%%%%%%%%%%%%%%%%%%%%%%%%%%%
%%%%%%%%%%%%%%%%%%%%%%%%%%%%%%%%%%%%%%%%%%%%%%%%%%%%%%%%%%%%%%%%%%%%%%%%%%%%%%%%%%%%%%%%%%%%%%%%%%%%%%%%%%%%%%%%%%%%%%%%%%%%%%%%%%
%%%%%%%%%%%%%%%%%%%%%%%%%%%%%%%%%%%%%%%%%%%%%%%%%%%%%%%%%%%%%%%%%%%%%%%%%%%%%%%%%%%%%%%%%%%%%%%%%%%%%%%%%%%%%%%%%%%%%%%%%%%%%%%%%%

\section{FPU $\pi$-mode under periodic boundary conditions}\label{FPU_OPM_PBC_section}

Let us start by considering the famous 1-dimensional lattice of $N$ particles with nearest neighbor interactions governed by the FPU-$\beta$ Hamiltonian \cite{Fermi1955,Fermi1974}
\begin{equation}\label{FPU_Hamiltonian_PBC}
H(\mathbf{q},\mathbf{p})=\frac{1}{2}\sum_{j=1}^{N}p_{j}^{2}+\sum_{j=1}^{N}\biggl[\frac{1}{2}(q_{j+1}-q_{j})^2+\frac{1}{4}\beta(q_{j+1}-q_{j})^4\biggr]=E
\end{equation}
where $q_{j}$ is the displacement of the $j$th particle from its equilibrium position, $p_{j}$ is the corresponding momentum, $\beta$ is a real positive constant and $E$ is the constant energy of the system.

We first turn our attention to the well-known $\pi$-mode of Eq. (\ref{FPU_Hamiltonian_PBC}), which has been studied in detail in several publications \cite{Budinsky1983,Poggi1997,Cretegnyetal1998,Cafarella2004,Antonopoulos2006IJBC,Leo2010}. It is defined by
\begin{equation}\label{FPU_non_lin_mode_periodic_boundary_conditions_OPM}
q_{j}(t)=-q_{j+1}(t)\equiv q(t),\;j=1,\ldots,N
\end{equation}
and exists only when periodic boundary conditions are imposed
\begin{equation}\label{FPU_periodic_boundary_conditions_OPM}
q_{N+k}(t)=q_{k}(t),\;k=1,\ldots,N;\;\forall t
\end{equation}
with $N$ even (for a comprehensive review of NNMs and their connection with discrete symmetries of the Hamiltonian, see \cite{Bountis2010}).

Our aim is to investigate here a chaotic region near this orbit at energies where it has just become unstable. Thus, we choose as our observable the quantity
\begin{equation}
\eta(t)=q_{\frac{N}{2}}(t)+q_{\frac{N}{2}-1}(t)
\end{equation}
which satisfies $\eta(t)=0$ exactly at the $\pi$-mode and remains close to zero at energies $E$ where the $\pi$-mode is still stable. However, $\eta(t)$ will eventually deviate from zero at energies above the first destabilization threshold, i.e. $E>E_{\mbox{u}}^1$. To compare our results with those published recently in \cite{Leo2010}, we restrict ourselves to the case of $N=128$ and $\beta=1$ for which $E_{\mbox{u}}^1\approx 0.0257$ \cite{Antonopoulos2006IJBC} and take as our total energy $E=0.768$ (i.e. $\epsilon=E/N=0.006$), at which our $\pi$-mode is certainly unstable.

%##########################################
\begin{figure}[!ht]
\centering{
\includegraphics[width=6cm,height=5cm]{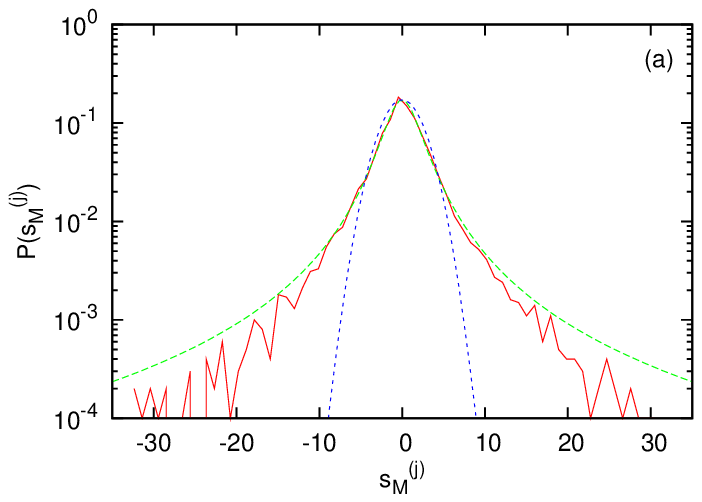}
\includegraphics[width=6cm,height=5cm]{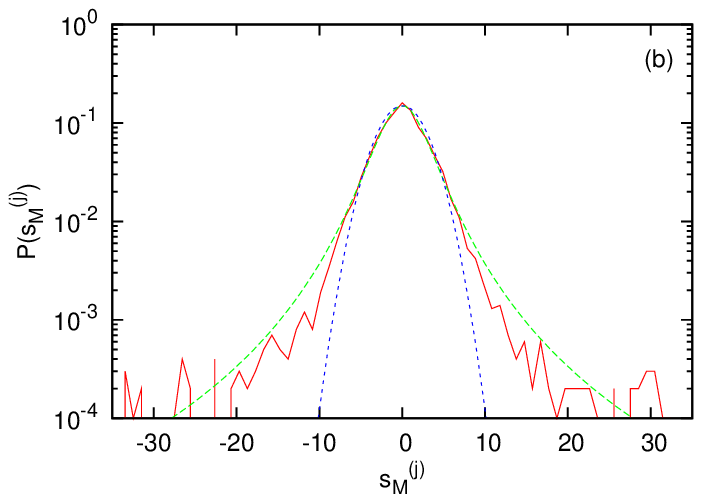}\\
\includegraphics[width=6cm,height=5cm]{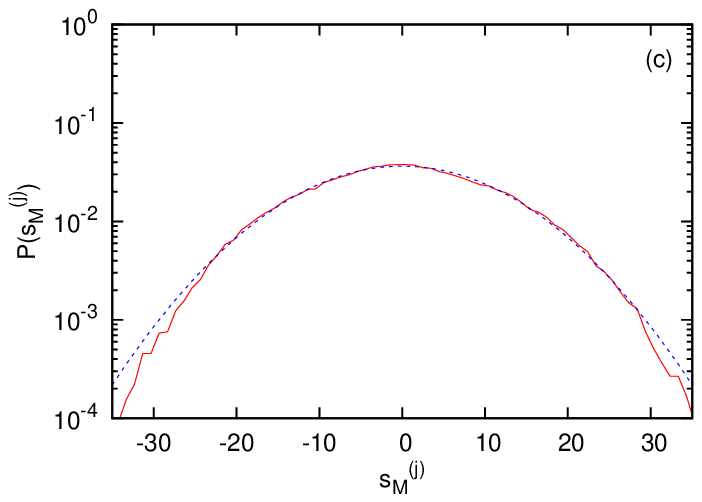}
\includegraphics[width=6cm,height=5cm]{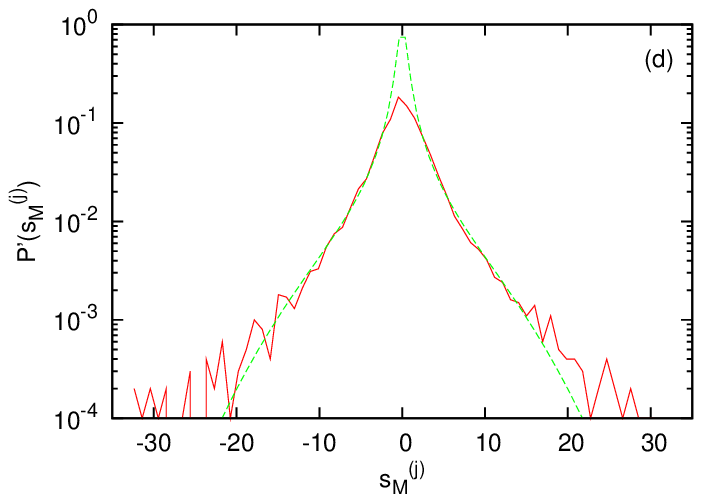}
\caption{Plot in linear-log scale of numerical (red curve), $q$-Gaussian (green curve) and Gaussian (blue curve) distributions for the FPU $\pi$-mode with periodic boundary conditions for $N=128$, $\beta=1$ and $\epsilon=0.006>\epsilon_{\mbox{u}}^{1}\approx0.0002$. Panel (a) corresponds to final integration time $t_{\mbox{f}}=10^5$ using  $N_{\mbox{ic}}=10^4$ time windows and $M=10$ terms in the computation of the sums. In this case the numerical fitting with a $q$-Gaussian gives $q\approx1.818$ with $\chi^2\approx0.0007$. Panel (b) corresponds to $t_{\mbox{f}}=10^6$, $N_{\mbox{ic}}=10^4$ and $M=100$. Here, the numerical fitting gives $q\approx1.531$ with $\chi^2\approx0.0004$. Panel (c) corresponds to $t_{\mbox{f}}=10^8$, $N_{\mbox{ic}}=10^5$ and $M=1000$. It is evident that the numerical distribution (red curve) has almost converged to a Gaussian (blue curve). Panel (d) shows the same red pdf as in panel (a) together with the $\tilde{P}$ function of Eq. (\ref{eq28paperTsallisTirnakli}) for $a_1\approx0.009$, $a_q\approx2.849$ and $q\approx2.179$ with $\chi^2\approx0.000\,08$ (green curve).}\label{fig_FPU_OPM_PBC_nrgdensity=0.006_CLT}}
\end{figure}
%##########################################

Regarding the accuracy of our integration, it is determined at each time step by requiring $|H(q(t),p(t))-E_{\mbox{in}}|<10^{-5}$, where $(q(t),p(t))$ is the numerical solution obtained using the Burlisch-Stoer (non-symplectic) integrator with an adaptive step size control \cite{Numericalrecipes} and $E_{\mbox{in}}$ is the initial energy at time $t=0$. We note that we have compared this method to Yoshida's 4th order symplectic integrator \cite{Yoshida1990} with time step equal to 0.05 (yielding a relative energy error of order $10^{-6}$ - $10^{-7}$) and have found nearly identical results in every case.

As we see in Fig. \ref{fig_FPU_OPM_PBC_nrgdensity=0.006_CLT}, when we increase the total integration time $t_{\mbox{f}}$, the pdfs (red curves) we obtain approach closer and closer to a Gaussian with $q$ tending to 1. Moreover, this seems to be independent of the values of $N_{\mbox{ic}}$ and/or $M$, at least up to the final integration time $t_{\mbox{f}}=10^8$ that we have been able to check. For example, when we vary in Fig. \ref{fig_FPU_OPM_PBC_nrgdensity=0.006_CLT}(b) the parameters $N_{\mbox{ic}}$ and $M$, we obtain $q$-Gaussians of very similar shape with $q$ between 1.51 and 1.67. It is important to note, however, that the same data may be better fitted by other similar looking functions: For example, we have found in the case of Fig. \ref{fig_FPU_OPM_PBC_nrgdensity=0.006_CLT}(a) that the numerical distribution (red curve) is more accurately approximated by a function presented in \cite{Tsallisbook2009,CelikogluTirnakli2010,TsallisTirnakli2010}
\begin{equation}\label{eq28paperTsallisTirnakli}
\tilde{P}(s_M^{(j)})=\frac{1}{\Bigr\{1-\frac{a_q}{a_1}+\frac{a_q}{a_1}\exp[(q-1)a_1s_M^{(j)2}]\Bigl\}^{\frac{1}{q-1}}},\;a_1,a_q\geq0\mbox{ and } q>1
\end{equation}
where $a_1\approx0.009,a_q\approx2.849$ and $q\approx2.179$ with $\chi^2\approx0.000\,08$, in contrast to the $\chi^2\approx0.0007$ obtained by fitting the same distribution by a $q$-Gaussian with $q\approx1.818$ (see Fig. \ref{fig_FPU_OPM_PBC_nrgdensity=0.006_CLT}(a)). Eq. (\ref{eq28paperTsallisTirnakli}) represents the crossover between $q$-Gaussians and Gaussians and takes into account finite size (and time) effects reflected in the lowering of the tails of our distributions.

Thus, we conclude that the chaotic state we have obtained is a QSS. Indeed, when we increase the final integration time beyond $t_{\mbox{f}}\simeq4\times10^7$, we observe that the Lyapunov exponents monotonically increase and attain bigger values than those computed up to $t_{\mbox{f}}\simeq4\times10^7$ (see panel (b) of Fig. \ref{fig_C0_vs_time}). This may signify that the trajectories drift away from the neighborhood of the $\pi$-mode and enter a larger chaotic subspace of the energy manifold. Indeed, calculating a function $D(t)$ representing the euclidean distance of our orbit from the exact $\pi$-mode, we observed a sharp rise in the $D(t)$ values at $t\approx7.5\times10^8$.

%##########################################
\begin{figure}[!ht]
\centering{
\includegraphics[width=6cm,height=5cm]{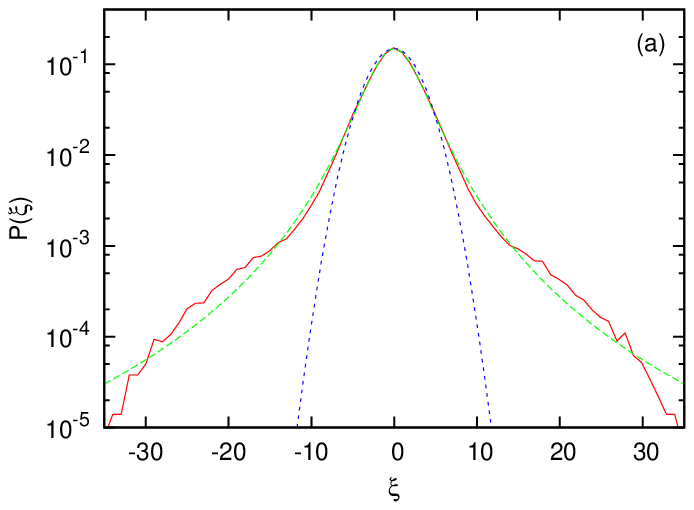}
\includegraphics[width=6cm,height=5cm]{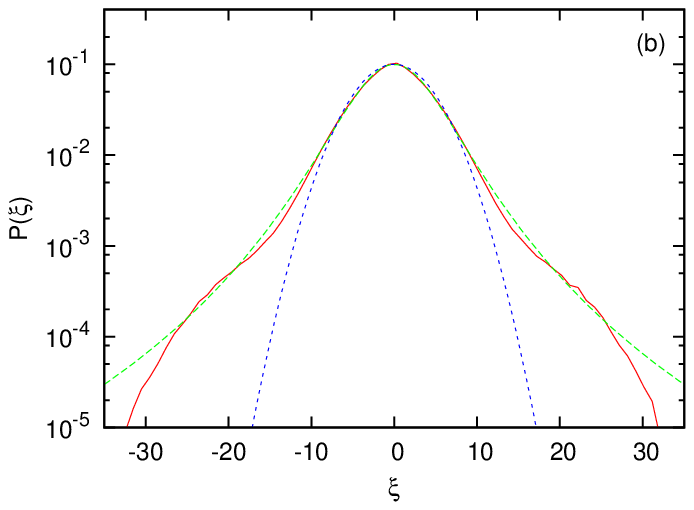}\\
\includegraphics[width=6cm,height=5cm]{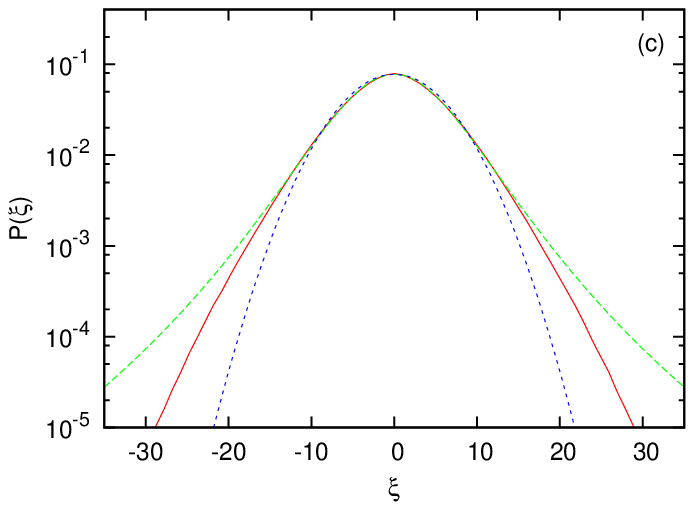}
\caption{Plot in linear-log scale of numerical (red curve), $q$-Gaussian (green curve) and Gaussian (blue curve) distributions for the FPU $\pi$-mode with periodic boundary conditions for $N=128$, $\beta=1$ and $\epsilon=0.006>\epsilon_{\mbox{u}}^{1}\approx0.0002$ following \cite{Leo2010}. $\xi$ is defined as in \cite{Leo2010} as well. Panel (a) corresponds to a final integration  time of $t_{\mbox{f}}=10^6$. In this case the numerical fitting with a $q$-Gaussian gives $q\approx1.495$ with $\chi^2\approx2.63\times 10^{-5}$. Panel (b) corresponds to $t_{\mbox{f}}=10^7$. Here, an analogous fitting gives $q\approx1.374$ with $\chi^2\approx3.86\times 10^{-5}$. Panel (c) corresponds to $t_{\mbox{f}}=10^8$ getting $q\approx1.279$ with $\chi^2\approx9.86\times 10^{-6}$.}\label{fig_FPU_OPM_PBC_nrgdensity=0.006_Tempesta}}
\end{figure}
%##########################################

Similar results concerning the tendency of the distributions to approach a Gaussian for $t_{\mbox{f}}>10^6$ are also obtained using the methodology proposed recently in \cite{Leo2010}. In this paper, however, the authors use a different way of calculating the numerical distributions of the $\eta_i$ observables: Instead of using sums, they divide their total time interval $[0,t_{\mbox{f}}]$ into short segments of 100 integration steps each and study the pdfs of their observables evaluated at the right endpoints of these segments. Moreover, since the longest integration time they considered was $t_{\mbox{f}}=10^6$, they only observed QSS approximated by $q$-Gaussians and did not detect their transitory character, which becomes evident for longer times. Indeed, when one takes $t_{\mbox{f}}=10^7$ or $10^8$, one finds that distributions computed by the strategy of \cite{Leo2010}, show a tendency to approach a Gaussian by finding that $q\rightarrow1$ as we can see in Fig. \ref{fig_FPU_OPM_PBC_nrgdensity=0.006_Tempesta} of the present paper.

%%%%%%%%%%%%%%%%%%%%%%%%%%%%%%%%%%%%%%%%%%%%%%%%%%%%%%%%%%%%%%%%%%%%%%%%%%%%%%%%%%%%%%%%%%%%%%%%%%%%%%%%%%%%%%%%%%%%%%%%%%%%%%%%%%
%%%%%%%%%%%%%%%%%%%%%%%%%%%%%%%%%%%%%%%%%%%%%%%%%%%%%%%%%%%%%%%%%%%%%%%%%%%%%%%%%%%%%%%%%%%%%%%%%%%%%%%%%%%%%%%%%%%%%%%%%%%%%%%%%%
%%%%%%%%%%%%%%%%%%%%%%%%%%%%%%%%%%%%%%%%%%%%%%%%%%%%%%%%%%%%%%%%%%%%%%%%%%%%%%%%%%%%%%%%%%%%%%%%%%%%%%%%%%%%%%%%%%%%%%%%%%%%%%%%%%

\section{Chaotic breathers and the FPU $\pi$-mode under Periodic Boundary Conditions}\label{OPM_chaotic_breathers}

%%%%%%%%%%%%%%%%%%%%%%%%%%%%%%%%%%%%%%%%%%%%%%%%%%%%%%%%%%%%%%%%%%%%%%%%%%%%%%%%%%%%%%%%%%%%%%%%%%%%%%%%%
%%%%%%%%%%%%%%%%%%%%%%%%%%%%%%%%%%%%%%%%%%%%%%%%%%%%%%%%%%%%%%%%%%%%%%%%%%%%%%%%%%%%%%%%%%%%%%%%%%%%%%%%%
%%%%%%%%%%%%%%%%%%%%%%%%%%%%%%%%%%%%%%%%%%%%%%%%%%%%%%%%%%%%%%%%%%%%%%%%%%%%%%%%%%%%%%%%%%%%%%%%%%%%%%%%%

In \cite{Cretegnyetal1998}, the authors studied the time evolution of an FPU-$\beta$ chain towards equipartition by considering initial conditions close to the $\pi$-mode under periodic boundary conditions. They observed that at energies well above the threshold where this mode becomes modulationally unstable, a remarkable localization phenomenon occurs. In particular, they report the spontaneous appearance of excitations that strongly resemble discrete breather solutions, but have a finite lifetime, while their dynamics is chaotic as indicated by their positive Lyapunov exponents. Moreover, their numerical results suggest that the lifetime of these chaotic breathers is related to the time necessary for the system to reach equipartition.

The appearance of chaotic breathers produced as time evolves can be monitored by evaluating the energy per site $1\leq n\leq N$ as
\begin{equation}\label{on_site_energy}
E_n=\frac{1}{2}p_n^2+\frac{1}{2}V(q_{n+1}-q_n)+\frac{1}{2}V(q_n-q_{n-1})
\end{equation}
where $V(x)=\frac{1}{2}x^2+\frac{\beta}{4}x^4$, thereby defining the function
\begin{equation}\label{C0}
C_0(t)=N\frac{\sum_{i=1}^{N}E_i^2}{(\sum_{i=1}^{N}E_i)^2}.
\end{equation}
Since, $C_0$ is of order one if $E_i=E/N$ at each site $i$ of the chain and of order $N$ if the energy is localized at only one site, this function can serve as an efficient indicator of energy localization in the chain.

In their experiments, Cretegny et al. used $N=128$, $\beta=0.1$, $E>E_u$, where $E_u$ is the lowest destabilization energy of the $\pi$-mode and plotted $C_0$ versus $t$. Distributing evenly the energy among all sites at $t=0$ according to the $\pi$-mode pattern, they observed that $C_0$ initially grows to relatively high values, indicating that the energy localizes at a few sites. After a certain time, however, $C_0$ reaches a maximum and decreases towards an analytically derived asymptotic value $\bar{C}_0\simeq1.795$ \cite{Cretegnyetal1998}, which is associated with the breakdown of the chaotic breather and the onset of energy equipartition in the chain.

\begin{figure}[!ht]
\centering{
\includegraphics[width=6cm,height=4.75cm]{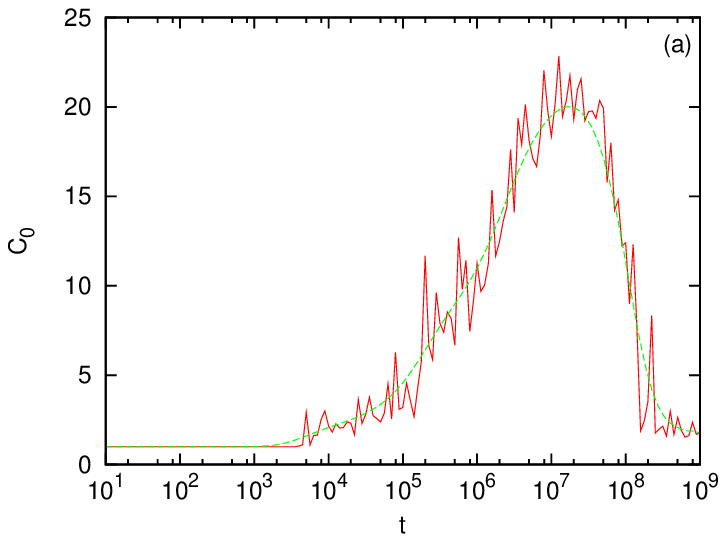}
\includegraphics[width=6cm,height=4.75cm]{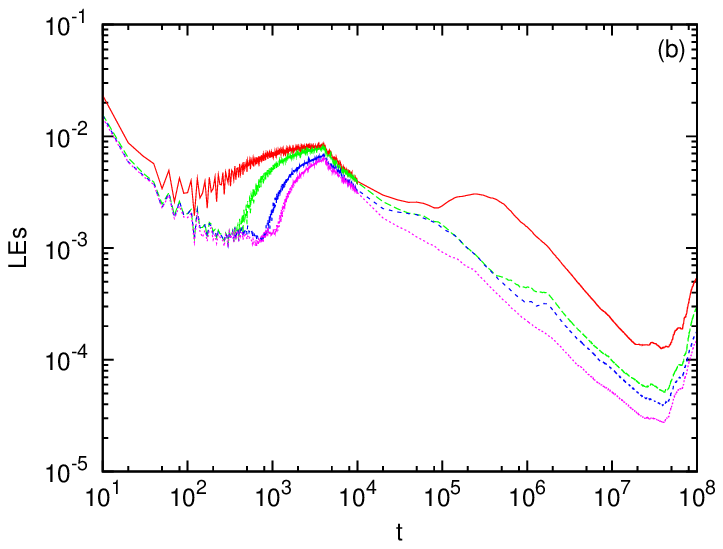}
\caption{Panel (a) is a plot of $C_0$ as a function of time for the unstable ($E=0.768>E_u\approx0.0257$) $\pi$-mode with $\beta=1$ and $N=128$. Panel (b) is a log-log plot of the four biggest Lyapunov exponents as a function of time for the same parameters as in panel (a). Repeating the calculation of the exponents for time windows of different size preserves the qualitative nature of the plot and changes only the smaller values of the exponents by one order of magnitude.}\label{fig_C0_vs_time}}
\end{figure}

We have performed the same study of the $\pi$-mode at a lower energy, $E=0.768>E_u\approx0.0257$, ($\beta=1$ and $N=128$ in the present work) and have been able to relate the $q$-Gaussian distributions of our QSS to the lifetime of chaotic breathers reported in \cite{Cretegnyetal1998} as follows: In panel (a) of Fig. \ref{fig_C0_vs_time}, we plot $C_0$ as a function of time and verify indeed that it grows over a long interval ($t\simeq1.8\times10^8$) after which it starts to decrease and eventually tends to the asymptotic value $\bar{C}_0\simeq1.795$ associated with energy equipartition. In panel (b) of the same figure we plot the evolution of the four biggest Lyapunov exponents and observe that initially they decrease towards zero until about $t\simeq4\times10^7$ when they start to increase towards positive values indicating the unstable character of the $\pi$-mode.

%##########################################
%##########################################

\begin{figure}[!ht] 
\centering{
\includegraphics[width=6cm,height=5cm]{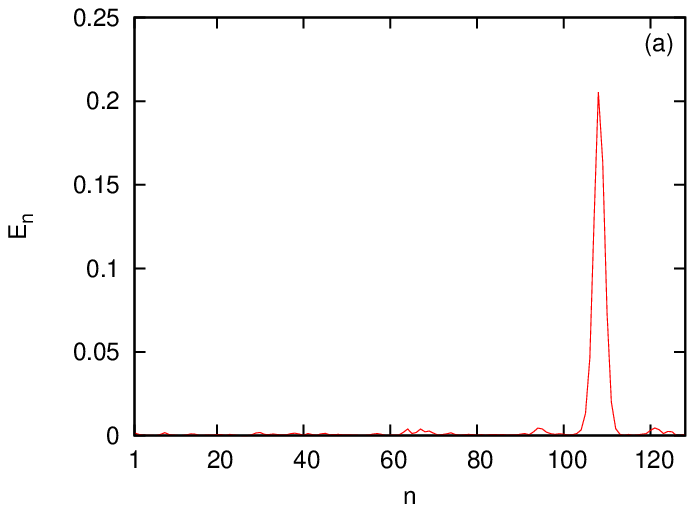}
\includegraphics[width=6cm,height=5cm]{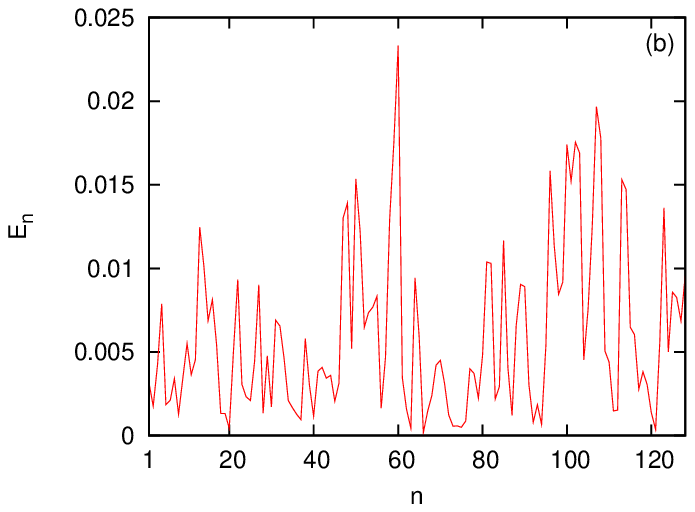}\\
\includegraphics[width=6cm,height=5cm]{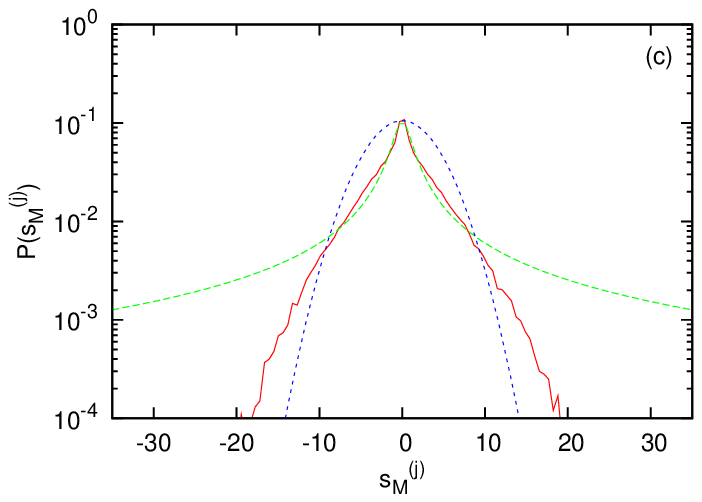}
\includegraphics[width=6cm,height=5cm]{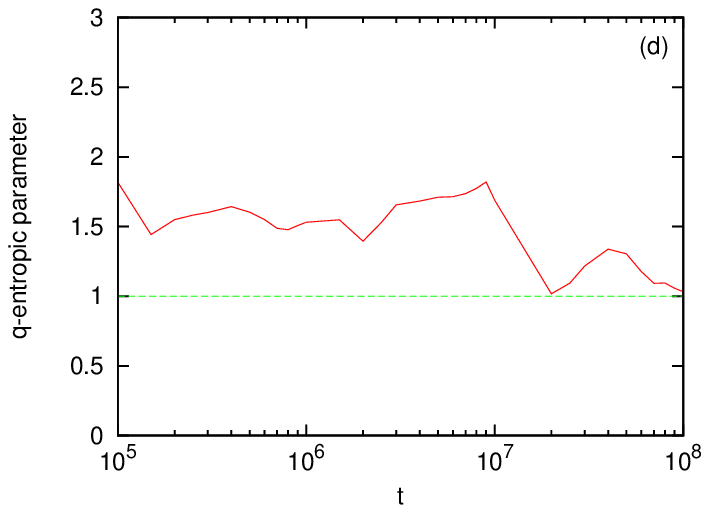}
\caption{In panel (a) at $t=10^7$, near the maximum of $C_0(t)$ (see Fig. \ref{fig_C0_vs_time}(a)), we see a chaotic breather. In panel (b) at $t=6\times10^8$, this breather has collapsed and the system has reached a state whose distribution is very close to a Gaussian, as implied already by the pdf shown in Fig. \ref{fig_FPU_OPM_PBC_nrgdensity=0.006_CLT}(c) at $t=10^8$. Note the scale difference in the vertical axes. Panel (c) at $t=10^7$ shows that the sum distribution, near the maximum of the chaotic breather, is still quite close to a $q$-Gaussian with $q\approx2.6$. Panel (d) presents an estimate of the $q$ index at different times, which shows that its values on the average fall significantly closer to 1 for $t>10^7$.}\label{fig_En_vs_time}}
\end{figure}

We now present in Fig. \ref{fig_En_vs_time} the instantaneous energies $E_n$ of all $N=128$ sites at two different times. In the first panel, at time $t=10^7$, where $C_0(t)$ attains a maximum (see Fig. \ref{fig_C0_vs_time}(a)), the energy  is clearly localized at very few sites, demonstrating the occurrence of a chaotic breather. At later times, however, e.g. $t=6\times10^8$, panel (b) shows that this breather has collapsed, as the system tends to equipartition. This is consistent with the fact that the pdf describing the statistics at $t=10^7$ is well-approximated by a $q$-Gaussian, with $q\approx2.6$, shown in panel (c), while at $t=10^8$, as we know from Fig. \ref{fig_FPU_OPM_PBC_nrgdensity=0.006_CLT}(c), this pdf is already very close to a true Gaussian. 

%##########################################
%##########################################

Finally, in Fig. \ref{fig_En_vs_time}(d) we present estimates of the $q$ values one obtains, when computing chaotic orbits near the $\pi$-mode at this energy density $\epsilon=E/N=0.006$. Remarkably enough, even though these values have an error bar of about $\pm10$ percent due to the different statistical parameters ($M,N_{\mbox{ic}}$) used in the computation, they exhibit a clear tendency to fall closer to 1 for $t>10^7$, where energy equipartition is expected to occur.

It is indeed a hard and open problem to determine exactly how equipartition times $T_{\mbox{eq}}$ scale with the energy density $\epsilon=E/N$ and other parameters (like $\beta$), particularly in the thermodynamic limit, even in 1-dimensional FPU lattices. Although it is a question that has long been studied in the literature \cite{DeLucaetal1995,Cretegnyetal1998,Berchiallaetal2004,Bambusietal2008,Benettinetal2009},
the precise scaling exponents by which $T_{\mbox{eq}}$ depends on $\epsilon,\;\beta$, etc. are not yet precisely known and the controversy is continuing to this very day.

\begin{figure}[!ht]
\centering{
\includegraphics[width=8cm,height=5cm]{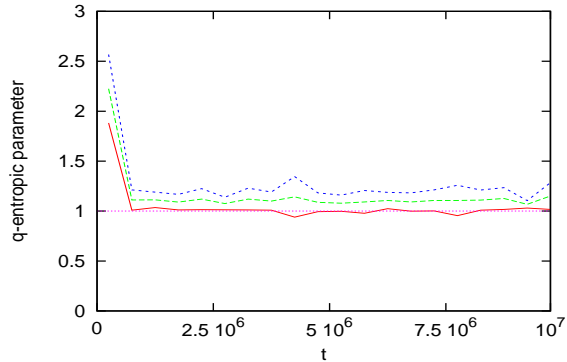}
\caption{Plot of $q$ values as a function of time for energy density $\epsilon=0.04$, i.e $E=5.12$ (green curve) for the $\pi$-mode of an FPU periodic chain with $N=128$ particles and $\beta=1$. The blue and red curves show error bars in the form of plus and minus one standard deviation. In agreement with other studies the transition to values close to $q=1$ occurs at $T_{\mbox{eq}} \approx7.5\times10^5$.} \label{q_function_time_E=5.12}}
\end{figure}

It would be very interesting if our $q$ values could help in this direction. In fact, carrying out more careful calculations, as in Fig. \ref{fig_En_vs_time}(d), at other values of the specific energy, e.g. $\epsilon=0.04$ (see Fig. \ref{q_function_time_E=5.12}) and $\epsilon=0.2$, we found $q$ plots that exhibited a clear decrease to values close to 1, at $T_{\mbox{eq}}\approx7.5\times10^5$ and $T_{\mbox{eq}}\approx7.5\times10^4$ respectively, approximately where the corresponding chaotic breathers collapse. Still, even though our results are consistent with what is known in the literature, the limited accuracy of our approach does not allow us to say something meaningful about scaling laws, as the system tends to equilibrium. In any case, the \textit{quantitative} usefulness of our $q$ estimates is a very interesting topic, which we are currently exploring and intend to describe in more details in a future publication.

In conclusion, comparing Figs. \ref{fig_FPU_OPM_PBC_nrgdensity=0.006_CLT}, \ref{fig_C0_vs_time}(a) and \ref{fig_En_vs_time}, we deduce that while the chaotic breather still exists, a QSS is observed fitted by $q$-Gaussians with $q$ well above unity, as seen in Fig. \ref{fig_En_vs_time}(c). However, as the chaotic breather breaks down for $t>10^8$ and energy equipartition is reached (see Fig. \ref{fig_En_vs_time}(b)), $q\rightarrow1$ and $q$-Gaussian distributions rapidly converge to Gaussians (see Fig. \ref{fig_FPU_OPM_PBC_nrgdensity=0.006_CLT}(c)) in full agreement with what is expected from BG Statistical Mechanics. 

%##########################################

%%%%%%%%%%%%%%%%%%%%%%%%%%%%%%%%%%%%%%%%%%%%%%%%%%%%%%%%%%%%%%%%%%%%%%%%%%%%%%%%%%%%%%%%%%%%%%%%%%%%%%%%%

\section{FPU SPO1 and SPO2 modes under fixed boundary conditions}\label{FPU_SPO1_FBC_section}

Let us now pass to the next part of our study and examine the chaotic dynamics near NNMs of the FPU system under \textit{fixed} boundary conditions. In particular, we first study pdfs of sums of chaotic orbit components near a NNM we call the SPO1 mode (see \cite{Ooyama1969,Antonopoulos2006IJBC}), which keeps one particle fixed for every two adjacent particles oscillating with opposite phase.

More specifically, we consider again the FPU-$\beta$ 1-dimensional lattice described by the Hamiltonian (\ref{FPU_Hamiltonian_PBC}) and impose fixed boundary conditions
\begin{equation}\label{FPU_fixed_boundary_conditions_SPO1}
q_{0}(t)=q_{N+1}(t)=0,\forall t
\end{equation}
with $N$ odd. The SPO1 mode is defined by
\begin{equation}\label{FPU_non_lin_mode_fixed_boundary_conditions_SPO1}
q_{2j}(t)=0,\qquad q_{2j-1}(t)=-q_{2j+1}(t),\;j=1,\ldots,\frac{N-1}{2}.
\end{equation}

\begin{figure}[!ht]
\centering{
\includegraphics[width=8cm,height=7cm]{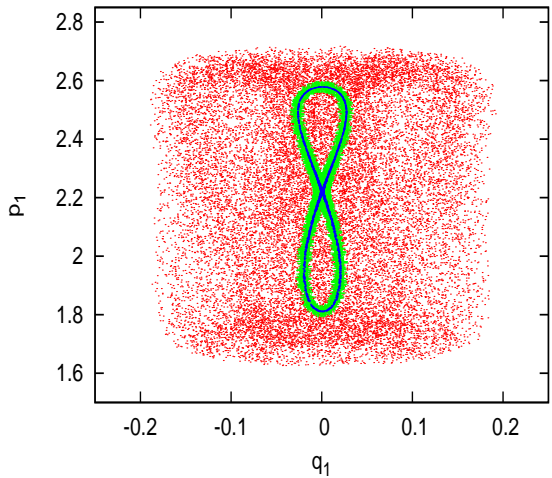}
\caption{The ``figure eight'' chaotic region (blue colour) for an initial condition at a distance about $1.192\times10^{-7}$ from the unstable SPO1 mode (depicted as the saddle point when $q_3=0$ and $p_3>0$), a slightly more extended ``figure eight'' region (green points) for an initial condition a little further away ($\approx1.086\times10^{-2}$) and a large scale chaotic region (red points) for an initial condition even more distant ($\approx3.421\times10^{-1}$) on the surface of section ($q_1,p_1$) computed at times when $q_3=0$. We have integrated our orbits up to $t_{\mbox{f}}=10^5$ using $E=7.4$, $N=5$ and $\beta=1.04$.} \label{PSS_FPU_SPO1_FBC_tf=1e5}}
\end{figure}

As shown in Fig. \ref{PSS_FPU_SPO1_FBC_tf=1e5}, the chaotic region very close to this solution (when it has just become unstable) appears for a long time isolated in phase space from other chaotic domains. In fact, one finds several such domains, embedded one into each other. For example, in the case $N=5$ with $\beta=1.04$ a ``figure eight'' chaotic region appears in blue traced by an orbit starting at a distance about $1.192\times10^{-7}$ from the SPO1 mode. This is easily seen on the surface of section ($q_1,p_1$) of Fig. \ref{PSS_FPU_SPO1_FBC_tf=1e5} computed at times when $q_3=0$ (and $p_3>0$) at the energy $E=7.4$. Even though the SPO1 mode is unstable (depicted as the saddle point for $q_3=0$) orbits starting sufficiently nearby remain in its vicinity for very long times, forming eventually the thin ``figure eight'' at the center of the figure.

Starting, however, at points a little further away (e.g. at an initial condition located at a distance about $1.086\times10^{-2}$ from the SPO1 mode), a more extended chaotic region is observed, plotted by green points, which still resembles a ``figure eight''. Choosing even more distant initial conditions (e.g. one that is located at a distance about $3.421\times10^{-1}$ from the SPO1 mode), a large scale chaotic region plotted by red points becomes evident in Fig. \ref{PSS_FPU_SPO1_FBC_tf=1e5}.

It is, therefore, reasonable to regard these different dynamical behaviours near the SPO1 mode as QSS and characterize them by pdfs of sum distributions, as explained in Section \ref{CLT_approach}. The idea behind this is that orbits starting initially at the immediate vicinity of the unstable SPO1 mode behave differently than those lying further away, since the latter orbits have the ability to explore more uniformly ergodic parts of the energy manifold. We thus expect to find, for the nearby regions of SPO1, $q$-Gaussian-like distributions with $1<q<3$, while for orbits starting sufficiently far the distributions are expected to be Gaussians with $q\rightarrow1$.

%##########################################
%##########################################

\begin{figure}[!ht]
\centering{
\includegraphics[width=3.5cm,height=3.5cm]{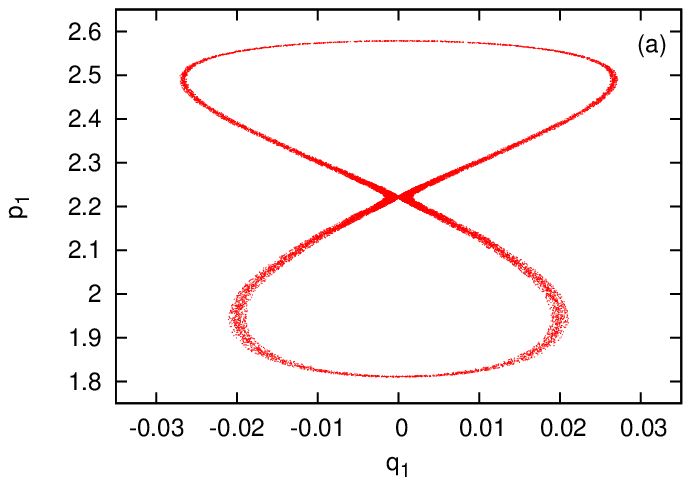}
\includegraphics[width=3.5cm,height=3.5cm]{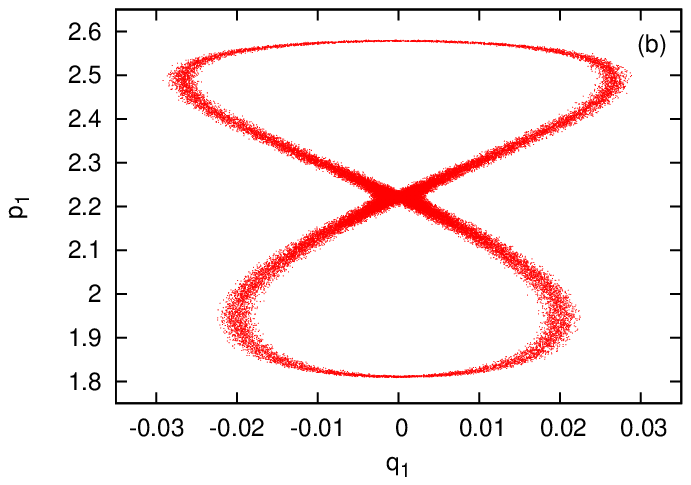}
\includegraphics[width=3.5cm,height=3.5cm]{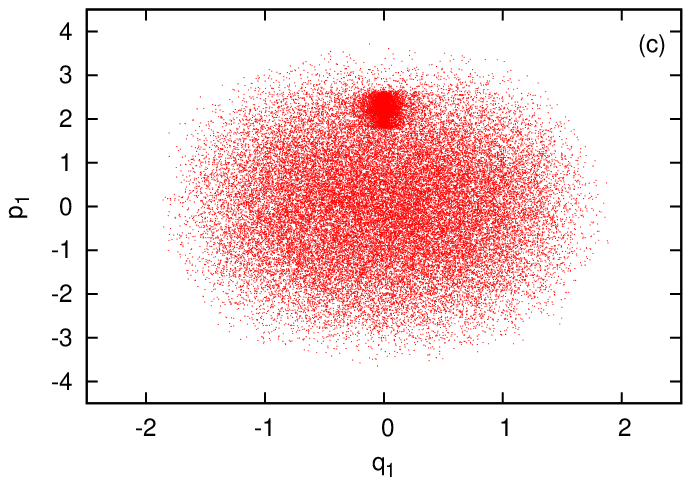}
\includegraphics[width=3.5cm,height=3.5cm]{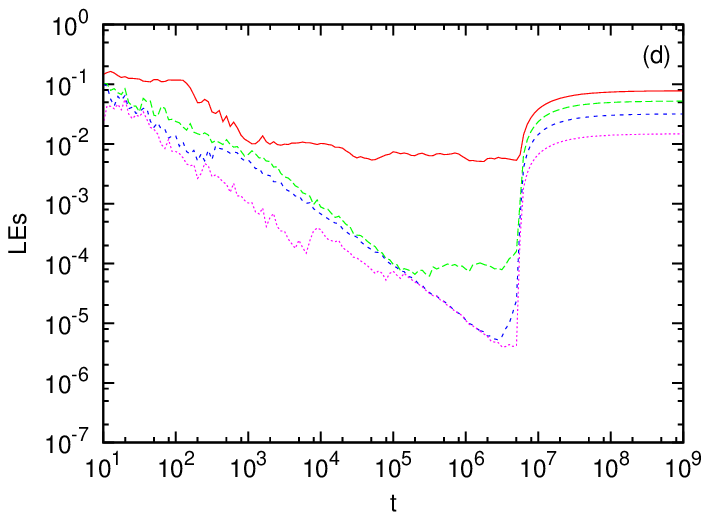}\\

\includegraphics[width=5cm,height=3.5cm]{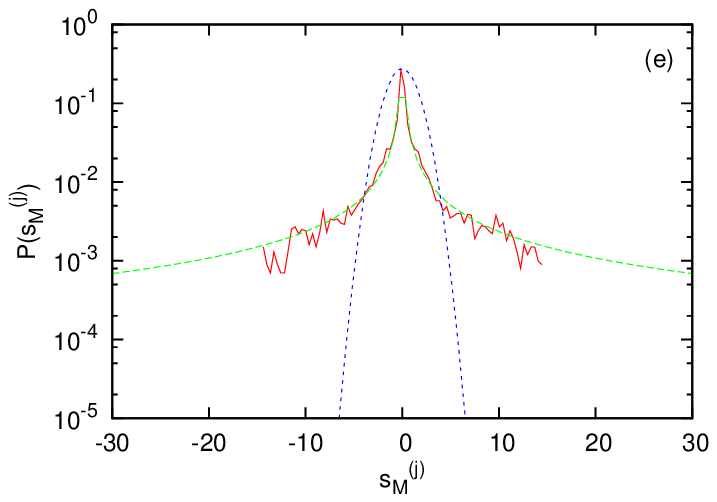}
\includegraphics[width=5cm,height=3.5cm]{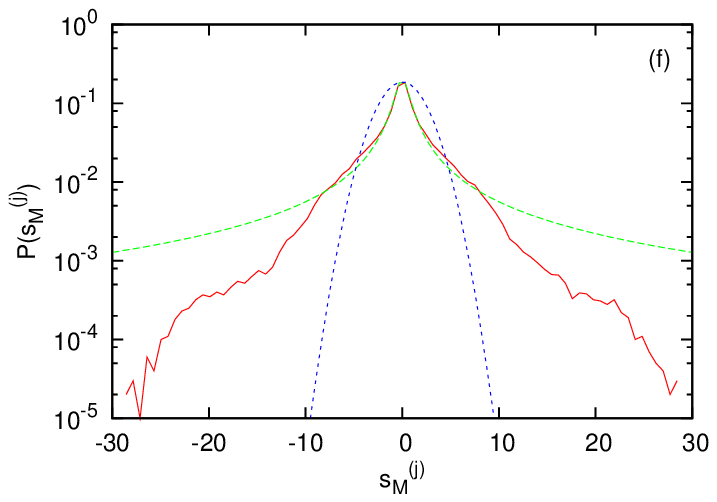}
\includegraphics[width=5cm,height=3.5cm]{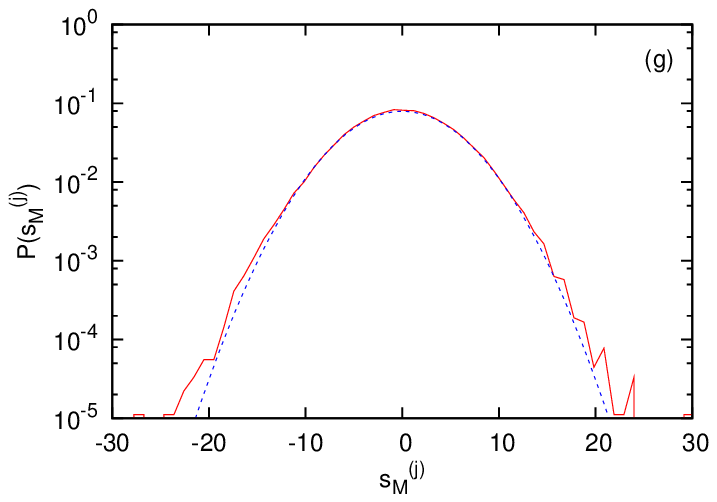}
\caption{Panel (a) shows the ($q_1,p_1$) surface of section of an orbit integrated up to $t_{\mbox{f}}=10^5$ and starting at a distance about $1.192\times10^{-7}$ from the unstable SPO1 for $N=5$ and $\beta=1.04$ at the energy $E=7.4$. Panel (b) is the same as panel (a) but for $t_{\mbox{f}}=10^7$. Panel (c) is the same as in panel (a) but for $t_{\mbox{f}}=10^8$. Panel (d) shows the corresponding four biggest Lyapunov exponents. Plots in linear-log scale of numerical (red curve), of $q$-Gaussian (green curve) and Gaussian (blue curve) distributions for the initial condition corresponding to panels (a) to (d). In panel (e) we use $t_{\mbox{f}}=10^5$, $N_{\mbox{ic}}=10^4$ and $M=10$. In this case the numerical fitting with a $q$-Gaussian gives $q\approx2.785$ with $\chi^2\approx0.000\,31$. In panel (f) integration time is $t_{\mbox{f}}=10^7$, $N_{\mbox{ic}}=10^5$ and use $M=1000$ terms. Here, the numerical fitting gives $q\approx2.483$ with $\chi^2\approx0.000\,47$. In panel (g) we set $t_{\mbox{f}}=10^8$, $N_{\mbox{ic}}=10^5$ and $M=1000$ and obtain a distribution much closer to a Gaussian ($q\approx1.05$), corresponding to the chaotic orbit shown in panel (c).} \label{fig_E=7.4_m=1.04_N=5_FPU_SPO1_FBC_dist=0.1192092895507812E-06_CLT}}
\end{figure}

To test the validity of these ideas, we have chosen the quantity
\begin{equation}
\eta(t)=q_{1}(t)+q_{3}(t)
\end{equation}
as our observable, which is exactly equal to zero at the SPO1 orbit (see Eq. (\ref{FPU_non_lin_mode_fixed_boundary_conditions_SPO1})). In fact, $\eta(t)$ remains very close to zero in the energy interval where the SPO1 mode is stable and becomes non-zero at energies beyond the first destabilization energy $E_{\mbox{u}}^1$ of the mode. Following what we have presented in Section \ref{CLT_approach}, we now study the three different initial conditions located in the neighborhood of the unstable SPO1 mode, seen in Fig. \ref{PSS_FPU_SPO1_FBC_tf=1e5}. In particular, in panel (a) of Fig. \ref{fig_E=7.4_m=1.04_N=5_FPU_SPO1_FBC_dist=0.1192092895507812E-06_CLT}, we see the surface of section created by the trajectory starting at a distance about $1.192\times10^{-7}$ from it and integrated up to $t_{\mbox{f}}=10^5$ while in the following two panels we see the same surface of section computed for final integration times of $t_{\mbox{f}}=10^7$ and $t_{\mbox{f}}=10^8$ respectively. The parameters are the same as in Fig. \ref{PSS_FPU_SPO1_FBC_tf=1e5}.

It is obvious that, as the integration time increases, even orbits which start close to the unstable SPO1 mode, eventually wander over a more extended part of phase space, covering gradually all of the energy manifold when the integration time is sufficiently large (e.g. $t_{\mbox{f}}=10^8$). This can also be explained by the Lyapunov exponents seen in panel (d). While for a fairly long time interval (up to approximately $t=10^7$) they decrease towards zero, they suddenly jump to much higher values after $t\approx10^7$, indicating that the orbit has entered into a wider chaotic domain of phase space.

\newpage
%##########################################
%##########################################
\newpage
An important question arises now: Are these different behaviours reflected in the statistical distributions associated with these trajectories computed for successively longer times? The answer to this question is presented in the three lower panels of Fig. \ref{fig_E=7.4_m=1.04_N=5_FPU_SPO1_FBC_dist=0.1192092895507812E-06_CLT}. In particular, in panel (e) we computed in linear-log scale the numerical (red curve), $q$-Gaussian (green curve) and Gaussian (blue curve) distributions for the initial condition located closest to the SPO1, using $t_{\mbox{f}}=10^5$, $N_{\mbox{ic}}=10^4$ time windows and $M=10$ terms in the computations of the sums. In this case, a fitting with the $q$-Gaussian (\ref{q_gaussian_distrib}) gives $q\approx2.785$ with $\chi^2\approx0.000\,31$. This distribution corresponds to the surface of section shown in panel (a). If we now increase $t_{\mbox{f}}$ by two orders of magnitude (see panel (f)) using $N_{\mbox{ic}}=10^5$, $M=1000$ and perform the same kind of fit we get $q\approx2.483$ with $\chi^2\approx0.000\,47$.

It is important to emphasize that the lower parts (tails) of the red distribution of panel (f) are not fitted well by the $q$-Gaussian. This suggests that by increasing the integration time, the initial $q$-Gaussian takes a transient form, which may very well be approximated by other types of functions like Eq. (\ref{eq28paperTsallisTirnakli}). This distribution corresponds to the surface of section of panel (b). By increasing the final integration time further to $t_{\mbox{f}}=10^8$ and using $N_{\mbox{ic}}=10^5$ and $M=1000$ terms, we observe that the red curve of panel (g) is very close to a Gaussian ($q\approx1.05$), characterizing the chaotic regime plotted in the surface of section of panel (c).

Let us now perform the same kind of study for the second initial condition of Fig. \ref{PSS_FPU_SPO1_FBC_tf=1e5}, starting at a distance about $1.086\times10^{-2}$ from the SPO1 periodic orbit. In this case we also obtain very similar results as in Fig. \ref{fig_E=7.4_m=1.04_N=5_FPU_SPO1_FBC_dist=0.1192092895507812E-06_CLT}. Again, the lower parts (tails) of the distribution deviate clearly from a $q$-Gaussian shape. In fact, for $t_{\mbox{f}}=10^8$, they are even closer to a Gaussian, as the corresponding orbit covers a much larger part of the available energy manifold.

%##########################################
\begin{figure}[!ht]
\centering{
\includegraphics[width=6cm,height=5cm]{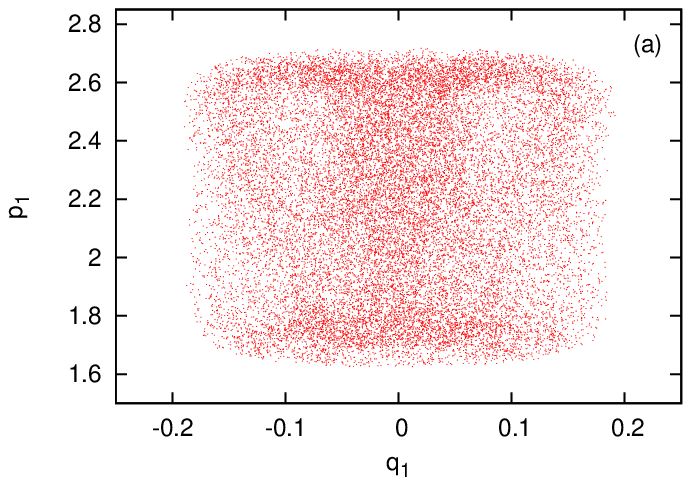}
\includegraphics[width=6cm,height=5cm]{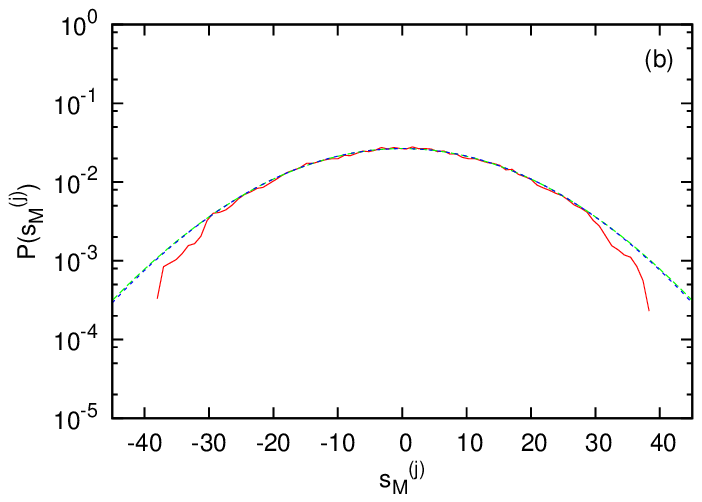}
\caption{Panel (a) is the ($q_1,p_1$) surface of section of an orbit in an initial distance about $3.421\times10^{-1}$ from the unstable SPO1 orbit integrated up to $t_{\mbox{f}}=10^5$. In panel (b) a plot of the data in linear-log scale shows that it is very well-fitted by a Gaussian (blue curve) with $q\approx1.0$. The final integration time is $t_{\mbox{f}}=10^5$ using $N_{\mbox{ic}}=10^5$ time windows and $M=1000$ terms in the computation of the sums.}\label{fig_E=7.4_m=1.04_N=5_FPU_SPO1_FBC_dist=0.3420654123880439E+00_CLT}}
\end{figure}
%##########################################

%##########################################
\begin{figure}[!ht]
\centering{
\includegraphics[width=4cm,height=5cm]{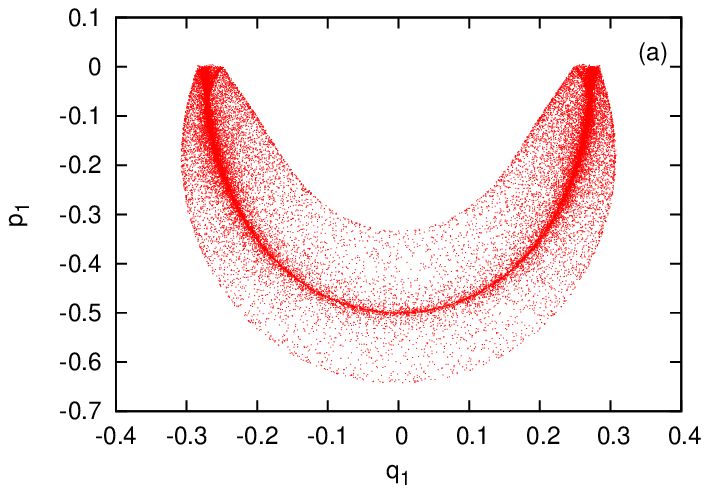}
\includegraphics[width=4cm,height=5cm]{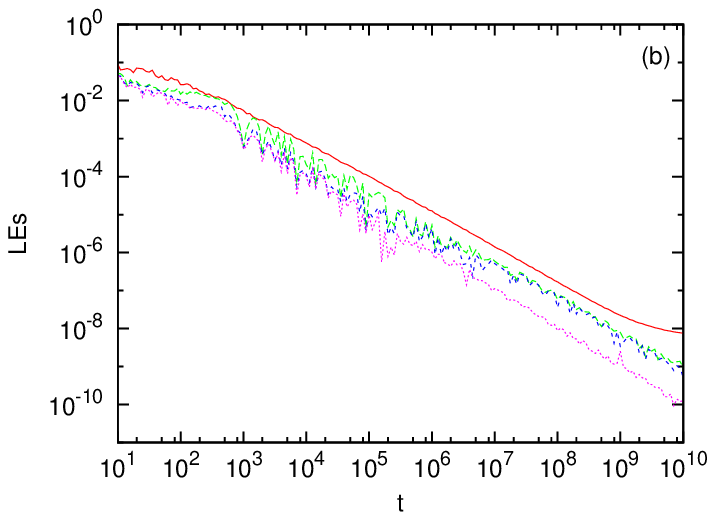}\\
\includegraphics[width=4cm,height=5cm]{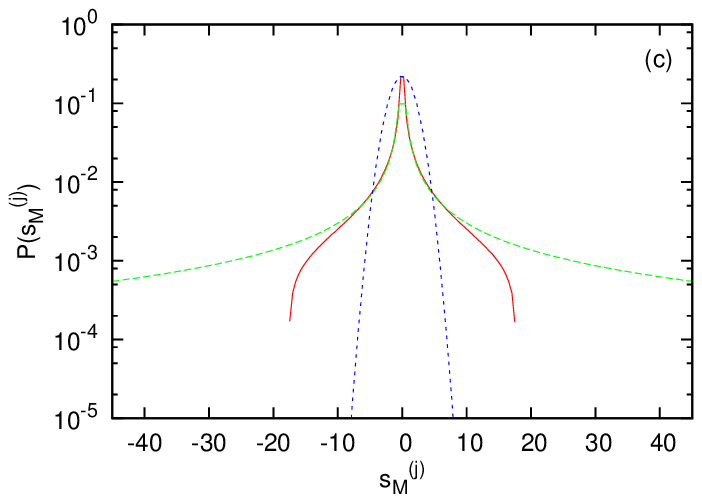}
\includegraphics[width=4cm,height=5cm]{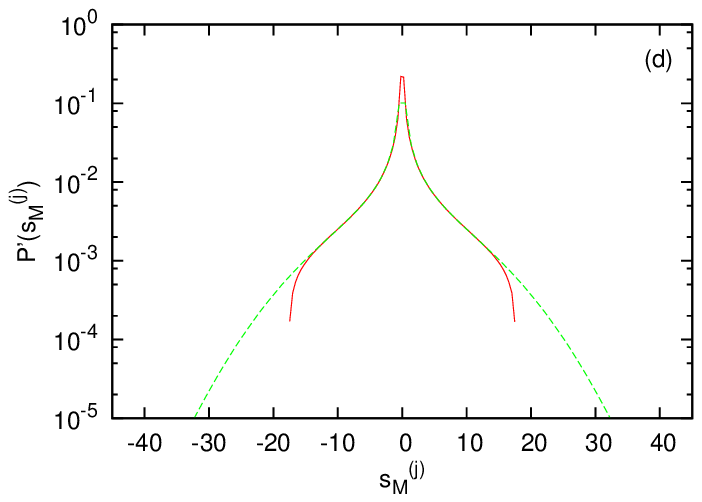}
\caption{Panel (a) is the ($q_1,p_1$) surface of section of an orbit integrated up to $t_{\mbox{f}}=10^{10}$ and starting initially at a distance about $1.418\times10^{-3}$ from the unstable SPO2 mode. Panel (b) shows the corresponding four biggest Lyapunov exponents. Panel (c) presents in linear-log scale the numerical (red curve), $q$-Gaussian (green curve) and Gaussian (blue curve) distributions. The final integration time is $t_{\mbox{f}}=10^{10}$ using $N_{\mbox{ic}}=333\,333\,333$ time windows and $M=10$ terms in the sums. Here, the numerical fitting with a $q$-Gaussian gives $q\approx2.769$ with $\chi^2\approx4.44\times 10^{-5}$, but panel (d) shows that the red distribution of panel (c) is better fitted by the $\tilde{P}$ function of Eq. (\ref{eq28paperTsallisTirnakli}) for $a_1\approx0.006$, $a_q\approx170$ and $q\approx2.82$ with $\chi^2\approx2.06\times10^{-6}$ (green curve).}\label{fig_E=0.5_m=1_N=5_FPU_SPO2_FBC_dist=0.1418209350721384E-02_CLT}}
\end{figure}
%##########################################

Finally, we have carried out the same analysis for the third initial condition of Fig. \ref{PSS_FPU_SPO1_FBC_tf=1e5} located at a distance about $3.421\times10^{-1}$ from the SPO1 mode. This corresponds to the orbit yielding the chaotic region shown in Fig. \ref{fig_E=7.4_m=1.04_N=5_FPU_SPO1_FBC_dist=0.3420654123880439E+00_CLT}(a). Note that despite the large size of this region compared to the ones shown in Fig. \ref{fig_E=7.4_m=1.04_N=5_FPU_SPO1_FBC_dist=0.1192092895507812E-06_CLT}(a) and (b) it is \textit{deeply embedded} in the chaotic domain of Fig. \ref{fig_E=7.4_m=1.04_N=5_FPU_SPO1_FBC_dist=0.1192092895507812E-06_CLT}(c). Due to the presence of ``strong chaos'' here, we find that even for small integration times ($t_{\mbox{f}}=10^5$), the distribution is very well approximated by the Gaussian shown in Fig. \ref{fig_E=7.4_m=1.04_N=5_FPU_SPO1_FBC_dist=0.3420654123880439E+00_CLT}(b).

In order to study further the persistence of QSS as time increases, we turned to another nonlinear mode of the FPU Hamiltonian with fixed boundary conditions called the SPO2 mode \cite{Antonopoulos2006PRE}. This is a NNM which keeps every third particle fixed, while the two in between move in exact out of phase motion. What is interesting about this NNM is that it becomes unstable at much lower energies (i.e. $E_{\mbox{u}}^1/N\propto N^{-2}$) compared to SPO1 ($E_{\mbox{u}}^1/N\propto N^{-1}$) \cite{Antonopoulos2006PRE}, much like the low $k=1,2,3,\ldots$ mode periodic orbits connected with the breakdown of FPU recurrences \cite{Christodoulidi2010,Bountis2010}. Thus, we expect that near SPO2, orbits will be more weakly chaotic than SPO1 and hence QSS are expected to persist for longer times. This is exactly what happens. As Fig. \ref{fig_E=0.5_m=1_N=5_FPU_SPO2_FBC_dist=0.1418209350721384E-02_CLT} clearly shows, the dynamics in a close vicinity of SPO2 has the features of what we might call ``edge of chaos'': Orbits wander in a regime of very small (positive) Lyapunov exponents, tracing a kind of ``banana'' shaped region much different than the ``figure eight'' we had observed near SPO1, while their pdfs (up to $t_{\mbox{f}}=10^{10}$), {\it converge} to a function that is close to a $q$-Gaussian, never deviating towards a Gaussian, as we have seen in the QSS of other FPU systems studied in this paper.

More specifically, let us take $N=5$, $\beta=1$, $E=0.5$ and choose an orbit located initially at a distance about $1.418\times10^{-3}$ from the SPO2 solution, which has just turned unstable (at $E_{\mbox{u}}^1\approx 0.4776$). As we can see in Fig. \ref{fig_E=0.5_m=1_N=5_FPU_SPO2_FBC_dist=0.1418209350721384E-02_CLT}(a), the dynamics here yields banana-like orbits at least up to $t_{\mbox{f}}=10^{10}$. The weakly chaotic nature of the motion in this domain is plainly depicted in panel (b) of this figure, where we have plotted the four positive Lyapunov exponents up to $t_{\mbox{f}}=10^{10}$. Note that, although they all decrease towards zero for a very long time, at about $t_{\mbox{f}}>10^9$, the largest one of them tends to converge to a very small value (about $10^{-8}$), indicating that the orbit is chaotically sticking to an ``edge of chaos'' region around SPO2.

In panel (c) of this figure, we have plotted the corresponding pdf at time $t_{\mbox{f}}=10^{10}$ (the function does not change after $t_{\mbox{f}}=10^7$). What we discover is an extremely long-lasting QSS, whose distribution is well-fitted by a $q$-Gaussian with $q\approx2.769$ and $\chi^2\approx4.44\times 10^{-5}$. The ``legs'' of the distribution away from the center deviate from the $q$-Gaussian shape, but remain very far from the Gaussian function also plotted in the figure by blue colour. We have also performed a similar fitting of our data with the function (\ref{eq28paperTsallisTirnakli}), as in the case of panel (d) of Fig. \ref{fig_FPU_OPM_PBC_nrgdensity=0.006_CLT}. What we find here is that the numerical distribution (red curve) of panel (c) is more accurately approximated by Eq. (\ref{eq28paperTsallisTirnakli}) where $a_1\approx0.006$, $a_q\approx170$ and $q\approx2.82$ with $\chi^2\approx2.06\times10^{-6}$, in contrast to the $\chi^2\approx4.44\times 10^{-5}$ obtained by fitting the same distribution by a $q$-Gaussian with $q\approx2.769$ (see Fig. \ref{fig_E=0.5_m=1_N=5_FPU_SPO2_FBC_dist=0.1418209350721384E-02_CLT}(a)).

Thus, we conclude that in sufficiently thin chaotic layers of multi-dimensional Hamiltonian systems with very small positive Lyapunov exponents it is possible to find non-Gaussian QSS that persist for quite long times as in the SPO2 case. Our numerical evidence suggests that in these regimes chaotic orbits stick for long time intervals to a complex network of islands, where their statistics is well approximated by distributions of the $q$-Gaussian type connected with Nonextensive Statistical Mechanics.

It is now interesting to study the dynamics near the unstable SPO1 and SPO2 modes using an analysis similar to the one carried out for the $\pi$-mode in Section \ref{OPM_chaotic_breathers}, following \cite{Cretegnyetal1998}. In particular, focusing on small perturbations of both modes under fixed boundary conditions, we wish to see how $C_0$ (see Eq. (\ref{C0})) displays the evolution towards energy equipartition and whether this transition will again be preceded by the appearance of chaotic breathers.

\begin{figure}[!ht]
\centering{
\includegraphics[width=6cm,height=4.75cm]{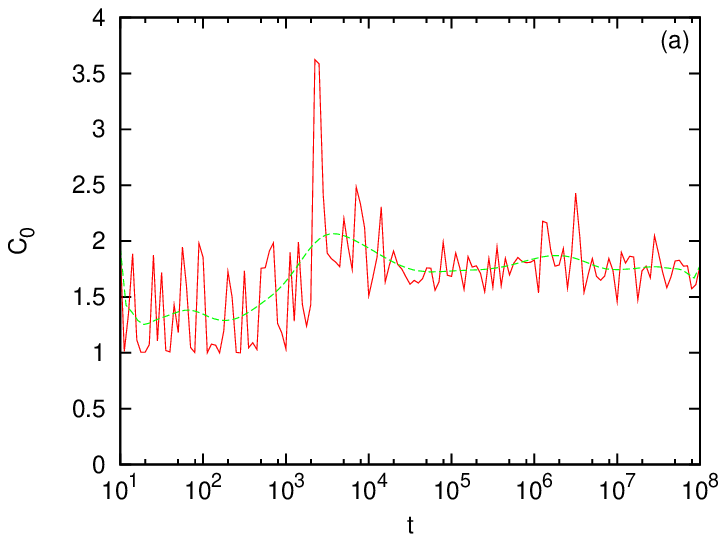}
\includegraphics[width=6cm,height=4.75cm]{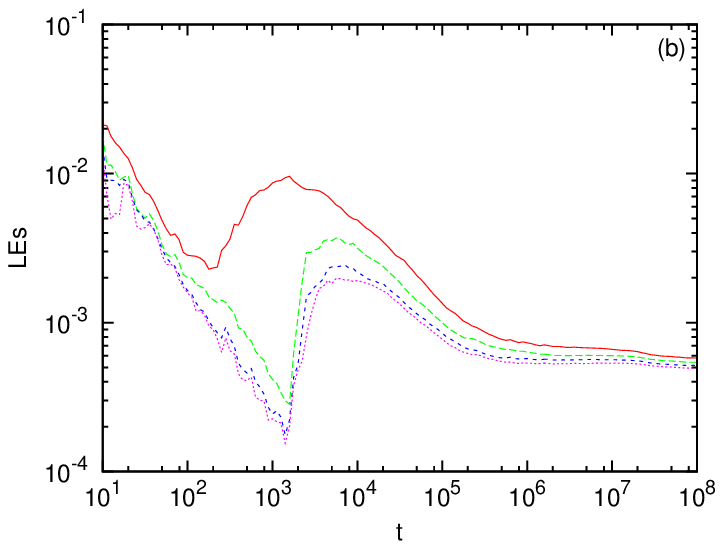}\\
\includegraphics[width=6cm,height=4.75cm]{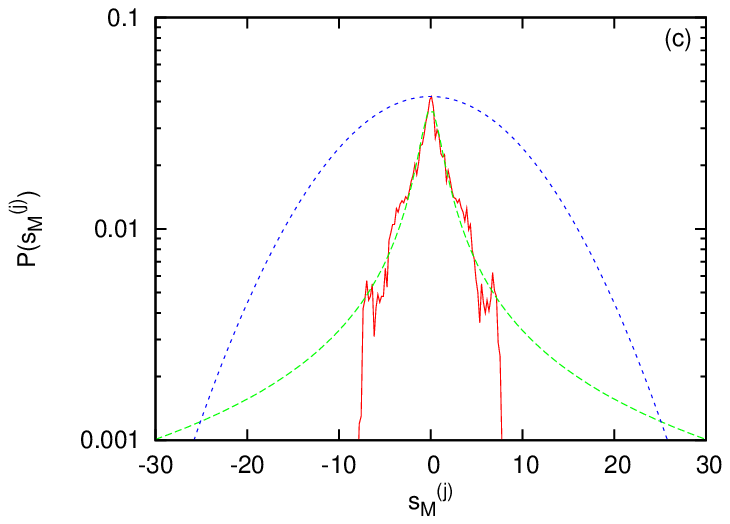}
\includegraphics[width=6cm,height=4.75cm]{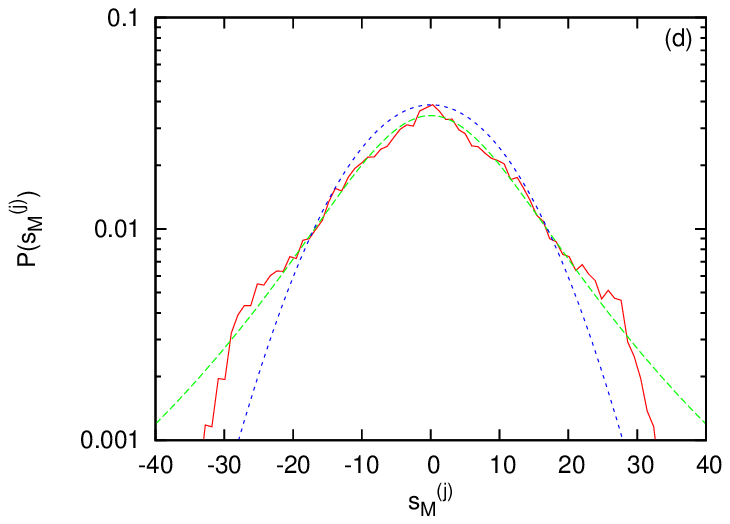}
\caption{Panel (a) is a plot of $C_0$ (red curve) as a function of time for a perturbation of the unstable ($E=1.5>E_u^{\mbox{SPO1}}\approx1.05226$) SPO1 mode with $\beta=1.04$ and $N=129$ at initial distance $2.22\times10^{-7}$. The green curve corresponds to the time average of the red curve. Panel (b) is a log-log scale plot of the four biggest Lyapunov exponents as a function of time. Plots in linear-log scale of numerical (red curve), $q$-Gaussian (green curve) and Gaussian (blue curve) distributions: In panel (c) for final integration time $t_{\mbox{f}}=4\times 10^5$, $N_{\mbox{ic}}=10^4$ time windows and $M=20$ terms in the sums, the numerical fitting with a $q$-Gaussian gives $q\approx2.843$ with $\chi^2\approx0.0003$, while panel (d) corresponds to $t_{\mbox{f}}=2\times10^6$, $N_{\mbox{ic}}=5\times10^4$ and $M=20$. Here, the numerical fitting gives $q\approx1.564$ with $\chi^2\approx0.00014$.}\label{fig_C0_vs_time_SPO1}}
\end{figure}

To find out, we used $N=129$ particles with $\beta=1.04$ and $E=1.5$ for the SPO1 mode and considered a neighboring orbit starting at an initial distance $2.22\times10^{-7}$ from it. For these parameters, $E_u^{\mbox{SPO1}}\approx1.05226$ \cite{Antonopoulos2006IJBC}. As we see in Fig. \ref{fig_C0_vs_time_SPO1}, $C_0$ grows on the average (without a clear maximum) and settles down to a value near $1.75$ indicating that the system reaches equipartition at about $t\approx10^6$, when the four biggest Lyapunov exponents begin to converge to their final values (see panel (b) of Fig. \ref{fig_C0_vs_time_SPO1}). Interestingly, during this period, the corresponding distribution in panel (c) (using the observable $\eta=q(64)+q(65)+q(66)$) is well fitted by a $q$-Gaussian with $q\approx2.843$ and $\chi^2\approx0.0003$ at $t=4\times 10^5$. For longer times, distributions quickly tend to Gaussians as shown in panel (d) for $t=2\times10^6$. Thus, even though (unlike the $\pi$-mode) no chaotic breather is observed here, during the time $C_0$ takes to relax to its limiting value, sum distributions are well approximated by $q$-Gaussians and only converge to Gaussians when energy equipartition has occurred.

%##########################################
%##########################################
 
\begin{figure}[!ht] 
\centering{  
\includegraphics[width=6cm,height=4.75cm]{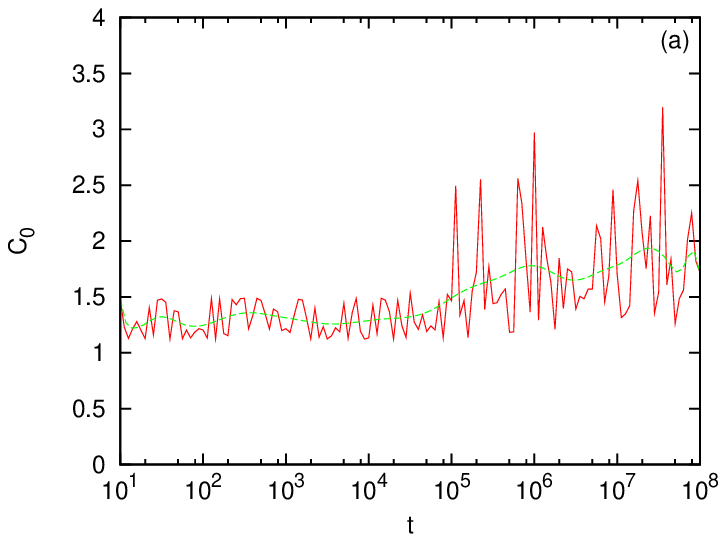}
\includegraphics[width=6cm,height=4.75cm]{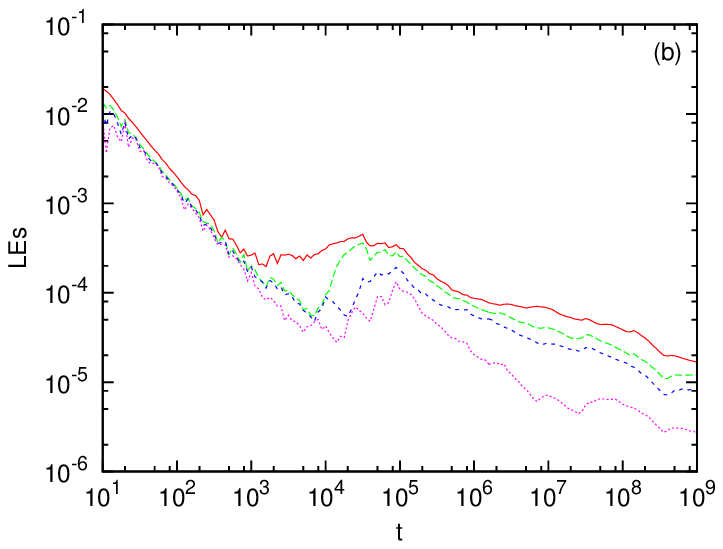}\\
\includegraphics[width=6cm,height=4.75cm]{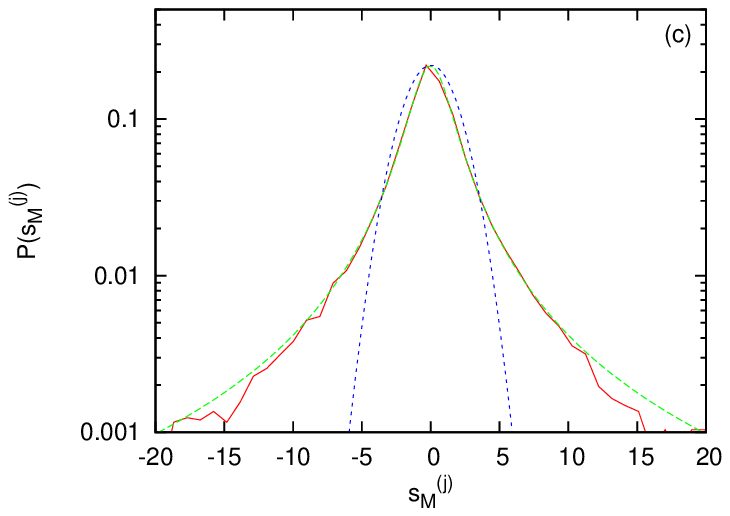}
\includegraphics[width=6cm,height=4.75cm]{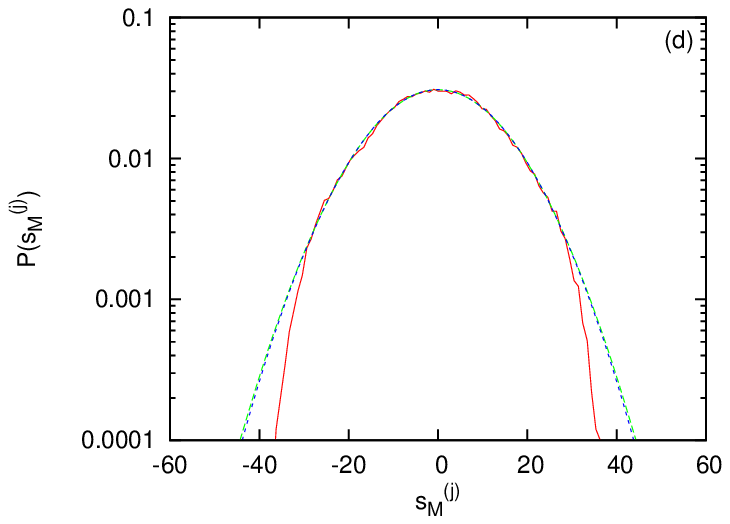}}
\caption{Panel (a) shows a plot of $C_0$ (red curve) as a function of time for a perturbation of the unstable ($E=0.1>E_u^{\mbox{SPO2}}\approx0.01279$) SPO2 mode with $\beta=1$ and $N=128$ at initial distance $4.09\times10^{-7}$. The green curve corresponds to the time average of the red curve. Panel (b) presents in log-log scale the four biggest Lyapunov exponents as a function of time for the same parameters and initial condition as in panel (a). In panel (c) we plot in linear-log scale of numerical (red curve), $q$-Gaussian (green curve) and Gaussian (blue curve) distributions for the same initial condition and parameters as in panel (a) for $t_{\mbox{f}}=10^6$, $N_{\mbox{ic}}=2.5\times10^4$ and $M=20$. Here, the numerical fitting gives $q\approx1.943$ with $\chi^2\approx0.00035$, which is further from a Gaussian than the corresponding pdf for SPO1 shown in Fig. \ref{fig_C0_vs_time_SPO1}(d). Panel (d) shows that the pdfs are much closer to Gaussian ($q\approx1.0$) at $t_{\mbox{f}}=10^8$, when the system has nearly reached the equipartition state.}\label{fig_C0_vs_time_SPO2}

\end{figure}

Let us repeat this process now for an orbit starting at an initial distance of $4.09\times10^{-7}$ from the SPO2 mode, with $N=128$, $\beta=1$ and $E=0.1$, above the first destabilization threshold of SPO2 (i.e. $E_u^{\mbox{SPO2}}\approx0.01279$) \cite{Antonopoulos2006PRE}. As we observe in Fig. \ref{fig_C0_vs_time_SPO2}, $C_0$ also does not exhibit a distinct high maximum here, but grows on the average \textit{more slowly} than in the SPO1 case. Thus, it takes longer to approach its limit, indicating that the system reaches equipartition at $t\approx4\times10^8$, where the four biggest Lyapunov exponents cease to decrease towards zero and tend to small positive values (see panel (b) of Fig. \ref{fig_C0_vs_time_SPO2}). The corresponding sum distribution shown in panel (c) at $t=10^6$ is still well fitted by a $q$-Gaussian with a high $q=1.943$ value, while we need to increase the time to $5\times10^8$ to see pdfs that are much closer to a Gaussian ($q\approx1.0$), as in panel (d), indicating that this is a lower bound for energy equipartition.

%##########################################
%##########################################
%%%%%%%%%%%%%%%%%%%%%%%%%%%%%%%%%%%%%%%%%%%%%%%%%%%%%%%%%%%%%%%%%%%%%%%%%%%%%%%%%%%%%%%%%%%%%%%%%%%%%%%%%%%%%%%%%%%%%%%%%%%%%%%%%%
%%%%%%%%%%%%%%%%%%%%%%%%%%%%%%%%%%%%%%%%%%%%%%%%%%%%%%%%%%%%%%%%%%%%%%%%%%%%%%%%%%%%%%%%%%%%%%%%%%%%%%%%%%%%%%%%%%%%%%%%%%%%%%%%%%
%%%%%%%%%%%%%%%%%%%%%%%%%%%%%%%%%%%%%%%%%%%%%%%%%%%%%%%%%%%%%%%%%%%%%%%%%%%%%%%%%%%%%%%%%%%%%%%%%%%%%%%%%%%%%%%%%%%%%%%%%%%%%%%%%%

\section{$q$-Gaussian distributions for a small microplasma system}\label{microplasma_section}

Let us turn our attention, finally, to a Hamiltonian system, which is very different than the FPU systems of the previous sections. In particular, we shall work with a system of few degrees of freedom characterized by \textit{long range interactions} of the Coulomb type. This will allow us to study statistically the dynamics of its transition from a crystalline-like to a liquid-like phase (the so-called ``melting transition'') \cite{Hill1994,Antonopoulos2010PRE} at small energies, as well as identify a transition from the liquid-like to a gas-like phase, as the total energy is constantly increased \cite{Gaspard2003}. Before presenting our results, we first describe the model and its various geometries.

Our microplasma system consists of $N$ ions of equal mass $m=1$ and electric charge $Q$ moving in a Penning trap with electrostatic potential \cite{Gaspard2003,Antonopoulos2010PRE}
\begin{equation}
\Phi(x,y,z)=V_{0}\frac{2z^{2}-x^{2}-y^{2}}{r_{0}^{2}+2z_{0}^{2}}
\end{equation}
and a constant magnetic field in the $z$ direction, whose vector potential is
\begin{equation}
\mathbf{A}(x,y,z)=\frac{1}{2}(-By,Bx,0).
\end{equation}
The system is described by the Hamiltonian
\begin{equation}
\mathcal{H}=\sum_{i=1}^{N}\Biggl\{\frac{1}{2m}[\mathbf{p}_{i}-q\mathbf{A}(\mathbf{r}_{i})]^{2}+Q\Phi(\mathbf{r}_{i})\Biggr\}+\sum_{1\leq
i<j\leq N}\frac{Q^{2}}{4\pi\epsilon_{0}r_{ij}}\label{initial_Hamiltonian}
\end{equation}
where $\mathbf{r_{i}}$ is the position of the $i$th ion, $r_{ij}$ is the Euclidean distance between $i$th and $j$th ions and $\epsilon_{0}$ is the vacuum permittivity. In the Penning trap, the ions are subjected to a harmonic confinement in the $z$ direction with frequency
\begin{equation}
\omega_{\mbox{z}}=\Biggl[\frac{4QV_{0}}{m(r_{0}^{2}+2z_{0}^{2})}\Biggr]^{\frac{1}{2}}
\end{equation}
while, in the perpendicular direction (due to the cyclotron motion), they rotate with frequency $\omega_{\mbox{c}}=QB/m$. Thus, in a frame rotating about the $z$ axis with Larmor frequency $\omega_{\mbox{L}}=\omega_{\mbox{c}}/2$, the ions are subjected to an overall harmonic potential with frequency $\omega_{x}=\omega_{y}=(\omega_{\mbox{c}}^{2}/4-\omega_{\mbox{z}}^{2}/2)^{1/2}$ in the direction perpendicular to the magnetic field. In the rescaled time $\tau=\omega_{\mbox{c}}t$, position $\mathbf{R}=\mathbf{r}/a$ and energy $H=\mathcal{H}/(m\omega_{\mbox{c}}^{2}a^{2})$ with $a=[Q^{2}/(4\pi\epsilon_{0}m\omega_{\mbox{c}}^{2})]^{1/3}$, the microplasma Hamiltonian (\ref{initial_Hamiltonian}) takes the form
\begin{equation}
H=\frac{1}{2}\sum_{i=1}^{N}\mathbf{P}_{i}^{2}+\sum_{i=1}^{N}\Bigl[\Bigl(\frac{1} {8}-\frac{\gamma^{2}}{4}\Bigr)(X_{i}^{2}+Y_{i}^{2})+\frac{\gamma^{2}}{2}Z_{i}^{2}\Bigr]+\sum_{i<j}\frac{1}{R_{ij}}=E\label{mic_plas_Ham}
\end{equation}
where $E$ is the total constant energy, $\mathbf{R}_{i}=(X_{i},Y_{i},Z_{i})$ and $\mathbf{P}_{i}=(P_{X_{i}},P_{Y_{i}},P_{Z_{i}})$ are the positions and momenta in $\mathbb{R}^3$ respectively of the $N$ ions, $R_{ij}$ is the Euclidean distance between different ions $i,j$ given by
\begin{equation}
R_{ij}=[(X_{i}-X_{j})^{2}+(Y_{i}-Y_{j})^{2}+(Z_{i}-Z_{j})^{2}]^{\frac{1}{2}}
\end{equation}
and finally $\gamma=\omega_{\mbox{z}}/\omega_{\mbox{c}}$.

The ions perform bounded motion under the condition that
\begin{equation}
0<|\gamma|<\frac{1}{\sqrt{2}}
\end{equation}
and the Penning trap is called prolate if $0<|\gamma|<1/\sqrt{6}$, isotropic if $|\gamma|=1/\sqrt{6}$ and oblate if $1/\sqrt{6}<|\gamma|<1/\sqrt{2}$. Thus, the motion is quasi 1-dimensional in the limit $\gamma\rightarrow0$ and quasi 2-dimensional in the limit $\gamma\rightarrow1/\sqrt{2}$. The $Z$ direction is a symmetry axis and hence, the $Z$ component of the angular momentum $L_{\mbox{Z}}=\sum_{i=1}^{N}X_{i}P_{Y_{i}}-Y_{i}P_{X_{i}}$ is conserved, being a second integral of the motion. We set, from here on, the angular momentum equal to zero (i.e. $L_{\mbox{Z}}=0$) and study the motion in the Larmor rotating frame.

In \cite{Antonopoulos2010PRE}, the authors demonstrate the occurrence of a dynamical regime change in a system (\ref{mic_plas_Ham}) composed of $N=5$ ions and confined in a prolate quasi 1-dimensional configuration of $\gamma=0.07$. More specifically, in the lower energy regime, a transition from crystalline-like to liquid-like behaviour is observed, called the ``melting phase''. It was first shown that this melting of the crystal is not associated with a sharp increase of the temperature at some critical energy, as might have been expected in first sight. Furthermore, the positive Lyapunov exponents are maximal at energies much higher than the ``melting'' regime. Thus, it appears that no clear macroscopic methodology is available for identifying and studying this melting process in detail.

For these reasons, the Smaller Alignment Index (SALI) method \cite{Skokos2001,Skokos2003,Skokos2004} was used by the authors to study the local microscopic dynamics in detail \cite{Antonopoulos2010PRE}. It was discovered that there exists an energy range of weakly chaotic behaviour, i.e. $E\in \Delta E_{\mbox{mt}}=(2,2.5)$, where the positive Lyapunov exponents are very small and the SALI exhibits a stair-like decay to zero with varying decay rates. This suggests the presence of long-lived ``sticky'' orbits executing a multi-stage diffusion process near the boundaries of resonance islands \cite{Skokos2008}. Thus, it was concluded that it is in this energy interval that the \textit{melting transition} of the system (\ref{mic_plas_Ham}) occurs.

Motivated by the above results and pursuing this investigation further, we decided to restrict ourselves to the same $\gamma$ and $N$, at which the minimum energy for ions to start moving appreciably about their equilibria positions is $E_0\approx1.8922$. Our first aim was to study the melting transition of the system as the energy of the Hamiltonian (\ref{mic_plas_Ham}) increases above $E_0$, using pdfs associated with chaotic trajectories to relate microscopic to macroscopic observables. Based on the results of the previous sections, we expected to find $q$-Gaussian approximations with $1<q<3$ in the vicinity of $\Delta E_{\mbox{mt}}$, where the positive Lyapunov exponents $\lambda_i,\;i=1,\ldots,3N$ are quite small compared to their maximum value of $\lambda_1\approx0.0558$ attained at $E\approx5.95$  \cite{Antonopoulos2010PRE}.

\begin{figure}[!ht]
\centering{
\includegraphics[width=6cm,height=5cm]{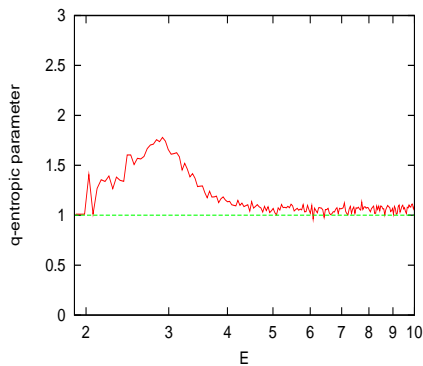}
\caption{Plot in log-linear scale of the $q$-entropic parameter (red colour) as a function of the energy $E$ of the microplasma (\ref{mic_plas_Ham}) for $\gamma=0.07$ (prolate trap) and $N=5$. In this plot, we have used $N_{\mbox{ic}}=2\times10^4$, $M=1000$ and have integrated every orbit up to $t_{\mbox{f}}=2\times10^7$. We have also plotted the line at $q=1$ (denoted by green) for comparison with the Gaussian case.}\label{fig_microplasma_gamma=0.07_N=5_CLT}}
\end{figure}

In Fig. \ref{fig_microplasma_gamma=0.07_N=5_CLT}, we show the results of our study for the interval $(E_0,10]$. We chose as an observable the quantity $\eta(t)=X_1(t)$, i.e. the first component of the position $\mathbf{R}_{1}$ of the first ion and set $N_{\mbox{ic}}=2\times10^4$, $M=1000$ and total integration time $t_{\mbox{f}}=2\times10^7$. Just as expected, in the energy range $\Delta E_{\mbox{mt}}$ of the melting transition, the values of the entropic parameter of the associated $q$-Gaussian pdf were well above $q=1$, indicating that the statistics is certainly \textit{not Gaussian}. In fact, the detected $q$-Gaussian shape persists up to $E\approx4$, implying perhaps that the energy interval of the melting transition is slightly bigger than originally estimated in \cite{Antonopoulos2010PRE}.

%##########################################
%##########################################

\begin{figure}[!ht]
\centering{
\includegraphics[width=6cm,height=5cm]{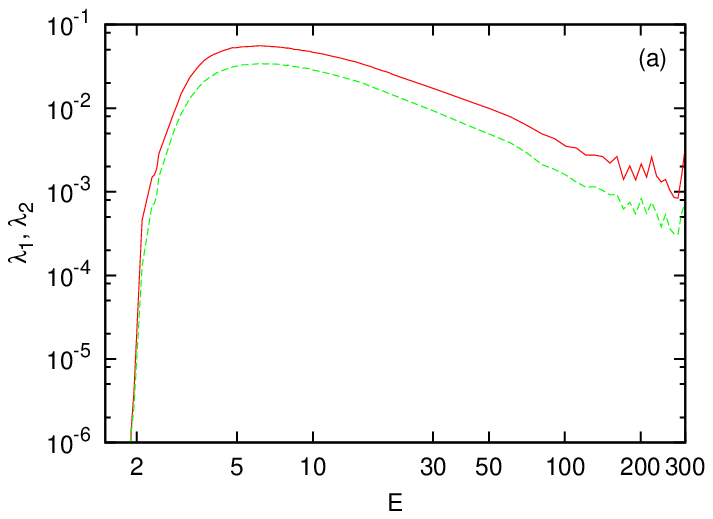}
\includegraphics[width=6cm,height=5cm]{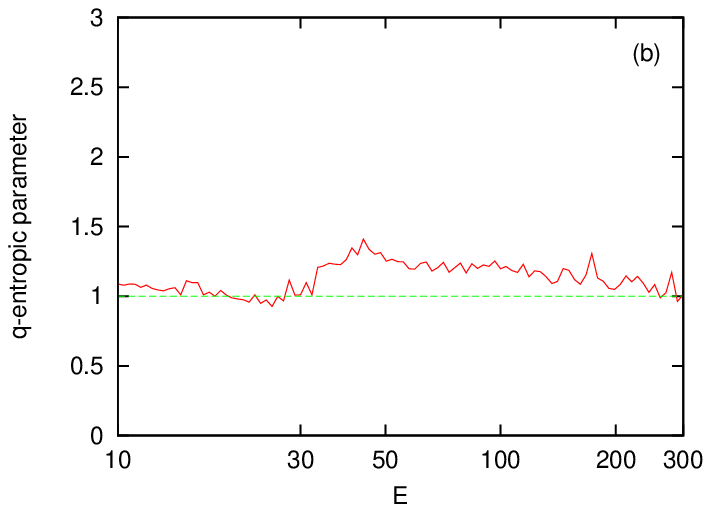}
\caption{Panel (a) shows the two biggest positive Lyapunov exponents ($\lambda_1$ with red and $\lambda_2$ with green colour) as a function of the energy $E$ in log-log scales. We observe that, after attaining their peak values at $E\approx5.8$, both of them  decay to zero according to formula (\ref{maxLE_prediction_high_nrgs}). Panel (b) is a plot in log-linear scales of the $q$-entropic parameter (red colour) as a function of the energy $E$ of the microplasma Hamiltonian (\ref{mic_plas_Ham}) for $\gamma=0.07$ and $N=5$. We have used $N_{\mbox{ic}}=2\times10^4$, $M=1000$, $t_{\mbox{f}}=2\times10^7$ and also plotted the line at $q=1$ (denoted by green) for reference to the entropic parameter of the Gaussian distribution.}\label{fig_microplasma_gamma=0.07_N=5_big_nrg_CLT}}
\end{figure}

Next, we proceeded to study the second dynamical state transition of system (\ref{mic_plas_Ham}), from a liquid-like to a gas-like regime, where the system evolves in a ``weakly chaotic'' way over an energy range where the largest positive Lyapunov exponents decrease towards zero following Eq. (\ref{maxLE_prediction_high_nrgs}), as shown in Fig. \ref{fig_microplasma_gamma=0.07_N=5_big_nrg_CLT}(a). We thus move to higher energies and concentrate on the passage of our system from a strongly chaotic regime at energies where the Lyapunov exponents attain their maximum (i.e. for $E\gtrsim6$) to energies where the motion is much less chaotic. In that regime, the highly chaotic but strongly correlated state of frequent inter-particle close encounters (``collisions'') is replaced by considerably fewer inter-particle energy/momenta exchanges and consequently much weaker chaos.

In \cite{Gaspard2003} it is demonstrated that if all $N^2$ inter-particle interaction terms of the Coulomb part of Hamiltonian (\ref{mic_plas_Ham}) at distances $R_{ij}$ contribute to the maximal Lyapunov exponent $\lambda_1$, one derives, for sufficiently large values of $T\rightarrow\infty$ (and therefore of $E\rightarrow\infty$), that
\begin{equation}\label{maxLE_prediction_high_nrgs}
\lambda_{1}\sim\left\langle \frac{N^{2}}{R_{ij}^{3}}\right\rangle \sim N\frac{(\ln T)^{\frac{1}{2}}}{T^{\frac{3}{4}}}
\end{equation}
where $T$ is related to the temperature of the system. This formula explains the asymptotic power law decay of the biggest Lyapunov exponent observed in panel (a) of Fig. \ref{fig_microplasma_gamma=0.07_N=5_big_nrg_CLT}. It can also be seen that the second largest Lyapunov exponent $\lambda_{2}$ (plotted with green colour) obeys a similar formula as a function of $E$.

It is, therefore, very interesting that the $q$ indices of the corresponding pdfs remain well above unity for $E\in(30,200)$, as shown in Fig. \ref{fig_microplasma_gamma=0.07_N=5_big_nrg_CLT}(b). This suggests that in this range there is a significant dynamical change of the system to a weaker form of chaos, associated with small positive Lyapunov exponents. Most probably, the $N^2$ inter-particle terms do not contribute significantly to the transition between the fully developed chaotic regime of correlated motion and the more ordered dynamics of a gas-like system. Thus, we conclude that the energy increase drives our system to a more and more organized state, favoring the kinetic over the Coulomb part, where very few close encounters between the particles occur, just like the case of an ``ideal gas''.

%##########################################
%##########################################

%%%%%%%%%%%%%%%%%%%%%%%%%%%%%%%%%%%%%%%%%%%%%%%%%%%%%%%%%%%%%%%%%%%%%%%%%%%%%%%%%%%%%%%%%%%%%%%%%%%%%%%%%%%%%%%%%%%%%%%%%%%%%%%%%%
%%%%%%%%%%%%%%%%%%%%%%%%%%%%%%%%%%%%%%%%%%%%%%%%%%%%%%%%%%%%%%%%%%%%%%%%%%%%%%%%%%%%%%%%%%%%%%%%%%%%%%%%%%%%%%%%%%%%%%%%%%%%%%%%%%
%%%%%%%%%%%%%%%%%%%%%%%%%%%%%%%%%%%%%%%%%%%%%%%%%%%%%%%%%%%%%%%%%%%%%%%%%%%%%%%%%%%%%%%%%%%%%%%%%%%%%%%%%%%%%%%%%%%%%%%%%%%%%%%%%%

\section{Conclusions}\label{conclusions}

In this paper, we have numerically constructed pdfs of rescaled sums of observables derived from chaotic trajectories of multi-dimensional Hamiltonian systems. Assuming that these observables behave as independent random variables, we sought to determine their statistics in phase space regions of ``weak chaos'' in the context of the Central Limit Theorem. More specifically, we focused on the neighborhood of unstable periodic orbits of various examples of FPU-$\beta$ particle chains and a microplasma system undergoing phase changes from crystalline to liquid and from liquid to gas-like dynamics, as its total energy increases.

For the FPU systems, we discovered that when chaotic orbits are restricted within ``small size'' domains near NNM which have just turned unstable, their pdfs are well approximated by $q$-Gaussian functions ($1<q<3$) for long times (typically up to $t\approx10^6$). These pdfs frequently represent QSS and tend to Gaussians ($q=1$) for longer integration times, as the orbits diffuse into domains of ``strong chaos'' characterized by large (positive) Lyapunov exponents. There are also cases, however, where the initial conditions are so close to an ``edge of chaos'' regime that the pdfs converge to a smooth function, which is well-approximated by different analytical expressions, valid in crossover regimes between $q$-Gaussians and Gaussians, as discovered recently also by other researchers in very different systems.

Thus, it is important to point out that $q$-Gaussians offer a \textit{family} of functions, parametrized by a single index $q$, that can successfully approximate sum distributions of QSS evolving in complicated networks of chaotic motion of multi-dimensional Hamiltonian systems. Indeed, $q$-Gaussians may represent the ``best'' choice for \textit{uniformly} fitting statistical data in cases where the passage from an ``edge of chaos'' regime to a domain of uniformly spread chaos is identified by a sequence of $q$ values starting well above 1 and tending to unity as time increases. With regard to the relation of $q$-Gaussians to certain important physical properties of FPU systems, such as energy equipartition, we have observed that these functions accurately represent long-lived transient states, whose $q$ is significantly larger than 1 and tend to Gaussians with $q$  falling quickly to unity, as the energy begins to be equally shared by all degrees of freedom.

In the case of the microplasma Hamiltonian, the approximation of our distributions by $q$-Gaussians allowed us to identify two distinct energy regimes of ``weakly'' chaotic behaviour (and small positive Lyapunov exponents): A low energy interval, $E\in[2,4]$, over which the system undergoes a melting transition and a much longer range, $E\in[30,200]$, where it passes from liquid-like to gas-like dynamics. This is accomplished  by observing over which energy domains the $q$-index of the distributions attains values significantly higher than unity, before returning to the $q=1$ value of Gaussians characterising ``strong chaos''.

Other authors have also studied QSS in conservative systems from the point of view of Nonextensive Statistical Mechanics \cite{Baldovin2004a,Baldovin2004b}, but do not compare sum distributions of their orbits, as we have done here. Rather, they calculate a quantity $T(t)$ called ``dynamical temperature'', which measures the total ``angular momentum'' of their system of $N$ coupled standard maps. They show that: (i) in the $N=1$ case, $T(t)$ grows smoothly to a $T_{\mbox{QSS}}(a)$ value, which depends on the nonlinearity parameter $a$ and is \textit{smaller}, for $a=O(1)$, than the $T_{\mbox{BG}}=1/12$ expected for a BG system, and (ii) for $N\geq2$-dimensional maps, $T(t)\rightarrow T_{\mbox{BG}}$ typically over $10^4$ iterations for $0.3<a<0.5$. Their initial conditions are chosen randomly within a thin layer that includes chaotic as well as regular orbits, thus it is not clear how to interpret their QSS and verify their results in more complicated systems. For example, the $T_{\mbox{QSS}}(a)<T_{\mbox{BG}}$ values for a single standard map may correspond to cases where most orbits remain within islands of regular motion.

On the other hand, in our study of FPU Hamiltonians, we have discovered QSS, which either diffuse to larger domains passing from a $q$-Gaussian to a Gaussian state, or may converge to a non-Gaussian type of pdf when the Lyapunov exponents are vanishingly small. It would be interesting, therefore, in a future publication, to define an analogous ``temperature'' quantity for our Hamiltonian systems, as was done in \cite{Baldovin2004a,Baldovin2004b} and study the times it needs to reach the BG state, as a function of the \textit{width} of the layer of initial conditions.

Interestingly, certain very recent results on low-dimensional area-preserving maps \cite{RuizBountisTsallis2010} are in very good agreement with what we have found here for multi-dimensional Hamiltonian systems. These results suggest that the presence of families of islands of stability in phase space and the associated stickiness phenomena in weakly chaotic regions of conservative systems are responsible for QSS approximated by $q$-Gaussian distributions. However, as soon as dissipative effects are included and strange attractors appear, chaotic motion becomes more uniform and sum distributions converge rapidly to Gaussians.

We believe, therefore, that the realm of conservative systems, represented by Hamiltonian flows or symplectic maps, is well suited for discovering complex metastable phenomena occurring in ``weakly chaotic'' domains. Even if the system is multi-dimensional, the dynamics near the boundaries of resonance islands and small positive values of Lyapunov exponents near these boundaries frequently yield long-lived states with nonextensive statistics. Thus, we suggest that it may be possible to apply these ideas to some very slow diffusive phenomena recently observed in 1-dimensional disordered chains \cite{Flach2009,Johansson2009,Skokos2009}, where it is not yet known if, in the limit of $t\rightarrow\infty$, diffusion will extend to particles further and further away or will be hindered by the boundary of some high-dimensional KAM torus, which is believed to exist \cite{Aubry2010}.

%%%%%%%%%%%%%%%%%%%%%%%%%%%%%%%%%%%%%%%%%%%%%%%%%%%%%%%%%%%%%%%%%%%%%%%%%%%%%%%%%%%%%%%%%%%%%%%%%%%%%%%%%%%%%%%%%%%%%%%%%%%%%%%%%%
%%%%%%%%%%%%%%%%%%%%%%%%%%%%%%%%%%%%%%%%%%%%%%%%%%%%%%%%%%%%%%%%%%%%%%%%%%%%%%%%%%%%%%%%%%%%%%%%%%%%%%%%%%%%%%%%%%%%%%%%%%%%%%%%%%
%%%%%%%%%%%%%%%%%%%%%%%%%%%%%%%%%%%%%%%%%%%%%%%%%%%%%%%%%%%%%%%%%%%%%%%%%%%%%%%%%%%%%%%%%%%%%%%%%%%%%%%%%%%%%%%%%%%%%%%%%%%%%%%%%%

\section*{Acknowledgments}

We are grateful to the referees for their useful comments and remarks. We also thank C. Tsallis, P. Tempesta and G. Ruiz-Lopez of the Centro Brasileiro de Pesquisas Fisicas at Rio de Janeiro for fruitful discussions during the preparation of the paper. Ch. A. was partially supported by the PAI 2007 - 2011 ``NOSY-Nonlinear Systems, Stochastic Processes and Statistical Mechanics'' (FD9024CU1341) contract of ULB and a grant from G.S.R.T., Greek Ministry of Education, for the project ``Complex Matter'', awarded under the auspices of the ERA Complexity Network. V. B. acknowledges the support of the European Space Agency under contract No. ESA AO-2004-070. T. B. is grateful for the hospitality of the Centro Brasileiro de Pesquisas Fisicas, at Rio de Janeiro, March 1 - April 5, 2010 and the Max Planck Institute for the Physics of Complex Systems in Dresden, April 5 - June 25, 2010, while work for this paper was carried out.

\section*{References}

\bibliographystyle{elsarticle-num}
\bibliography{paper_bibliography(new).bib}

\begin{thebibliography}{10}
\expandafter\ifx\csname url\endcsname\relax
  \def\url#1{\texttt{#1}}\fi
\expandafter\ifx\csname urlprefix\endcsname\relax\def\urlprefix{URL }\fi
\expandafter\ifx\csname href\endcsname\relax
  \def\href#1#2{#2} \def\path#1{#1}\fi

\bibitem{Anosov1967}
D.-V. Anosov, Geodesic flows on a compact {Riemann} manifold of negative
  curvature, Trudy Mat. Inst. Steklov 90.

\bibitem{Arnold1967}
V.-I. Arnold, A.~Avez, {Probl\`{e}mes} {Ergodiques} de la {M\'{e}canique}
  {Classique}, Gauthier-Villars, Paris, 1967 \& Benjamin, New York, 1968, 1967.

\bibitem{Sinai1972}
G.~Sinai, Y, Gibbs measures in ergodic theory, Uspekhi Matematicheskikh Nauk
  27~(4) (1972) 21.

\bibitem{Pesin1976}
B.~Pesin, Y, Invariant manifold families which correspond to nonvanishing
  characteristic exponents, Izv. Akad. Nauk. SSSR Ser. Mat. 40~(6) (1976) 1332.

\bibitem{Pesin1977}
B.~Pesin, Y, Lyapunov characteristic exponents and smooth ergodic theory,
  Uspekhi Matematicheskikh Nauk 32~(4) (1977) 196.

\bibitem{Ruelle1979}
D.~Ruelle, Ergodic theory of differentiable dynamical systems, Phys. Math. IHES
  50 (1979) 275.

\bibitem{katok1980}
A.~Katok, Liapunov exponents, entropy and periodic orbits for diffeomorphisms,
  Publications Math\'{e}matiques de l'IH\'{E}S 51 (1980) 137.

\bibitem{Ruelle1980}
D.~Ruelle, Measures describing a turbulent flow, Annals of the New York Academy
  of Sciences 357 (1980) 1.

\bibitem{Ruelle1982}
D.~Ruelle, Large volume limit of the distribution of characteristic exponents
  in turbulence, Communications in Mathematical Physics 87 (1982) 287.

\bibitem{Eckmann1985}
J.-P. Eckmann, D.~Ruelle, Ergodic theory of chaos and strange attractors, Rev.
  Mod. Phys. 57~(3) (1985) 617--656.

\bibitem{katok1985}
A.~Katok, J.-M. Strelcyn, F.~Ledrappier, F.~Przytycki, {Smooth} maps with
  singularities, invariant manifolds, entropy and billiards, no.~56,
  Universit\'{e} Paris-Nord, Pr\'{e}publications Math\'{e}matique, 1985.

\bibitem{Aizawa1984}
Y.~Aizawa, {Symbolic} dynamics approach to the two-dimensional chaos in
  area-preserving maps, Prog. Theor. Phys. 71 (1984) 1419--1421.

\bibitem{Chirikov1984}
B.-V. Chirikov, D.-L. Shepelyansky, {Correlation} properties of dynamical chaos
  in {Hamiltonian} diffusion, Physica D 13 (1984) 395--400.

\bibitem{Meiss1986}
J.-D. Meiss, E.~Ott, {Markov}-tree model transport in area preserving maps,
  Physica D 20 (1986) 387--402.

\bibitem{Skokos2008}
C.~Skokos, C.~Antonopoulos, T.-C. Bountis, Detecting chaos, determining the
  dimensions of tori and predicting slow diffusion in {Fermi}-{Pasta}-{Ulam}
  lattices by the {Generalized} {Alignment} {Index} method, European Physical
  Journal: Special Topics 165 (2008) 5--14.

\bibitem{Flach2009}
S.~Flach, D.-O. Krimer, C.~Skokos, Universal spreading of wavepackets in
  disordered nonlinear systems, Phys. Rev. Lett. 102 (2009) 024101.

\bibitem{Johansson2009}
M.~Johansson, G.~Kopidakis, S.~Lepri, S.~Aubry, Transmission thresholds in
  time-periodically driven nonlinear disordered systems, Europhysics Letters
  86~(1) (2009) 10009.

\bibitem{Skokos2009}
C.~Skokos, D.-O. Krimer, S.~Komineas, S.~Flach, Delocalization of wave packets
  in disordered nonlinear chains, Phys. Rev. E. 79 (2009) 056211.

\bibitem{Rice1995}
J.~Rice, {Mathematical} {Statistics} and {Data} {Analysis}, Duxbury Press
  (Second edition), 1995.

\bibitem{Tsallisbook2009}
C.~Tsallis, {Introduction} to {Nonextensive} {Statistical} {Mechanics}:
  {Approaching} a {Complex} {World}, Springer, New York, 2009.

\bibitem{Baldovin2004a}
F.~Baldovin, E.~Brigatti, C.~Tsallis, Quasistationary states in low-dimensional
  {Hamiltonian} systems, Phys. Lett. A 320 (2004) 254--260.

\bibitem{Baldovin2004b}
F.~Baldovin, M.-G. Moyano, A.-P. Majtey, A.~Robledo, C.~Tsallis, Ubiquity of
  metastable to stable crossover in weakly chaotic dynamical systems, Physica
  {\rm A} 340 (2004) 205--218.

\bibitem{CelikogluTirnakli2010}
A.~Celikoglu, U.~Tirnakli, S.~M. Duarte~Queiros, Analysis of return
  distributions in the coherent noise model, Phys. Rev. E. 82 (2010) 021124.

\bibitem{TsallisTirnakli2010}
C.~Tsallis, U.~Tirnakli, Nonadditive entropy and {Nonextensive Statistical
  Mechanics} - {Some} central concepts and recent applications, Journal of
  Physics: Conference Series 201 (2010) 012001.

\bibitem{Dauxois2007}
T.~Dauxois, {Non}-{Gaussian} distributions under scrutiny, J. Stat. Mech. DOI:
  10.1088/1742-5468/2007/08/N08001.

\bibitem{HilhorstSchehr2007}
J.~Hilhorst, H, G.~Schehr, A note on a {$q$-Gaussians} and {non-Gaussians} in
  {Statistical} {Mechanics}, J. Stat. Mech. DOI:
  10.1088/1742-5468/2010/06/P06003.

\bibitem{Hilhorst2010}
J.~Hilhorst, H, {Note} on a $q$-modified {Central} {Limit} {Theorem}, J. Stat.
  Mech. DOI: 10.1088/1742-5468/2010/10/P10023.

\bibitem{Budinsky1983}
N.~Budinsky, T.~Bountis, Stability of nonlinear modes and chaotic properties of
  $1$d {Fermi}-{Pasta}-{Ulam} lattices, Physica {\rm D} 8 (1983) 251--272.

\bibitem{Poggi1997}
P.~Poggi, S.~Ruffo, Exact solutions in the {FPU} oscillator chain, Physica {\rm
  D} 103 (1997) 251--272.

\bibitem{Dauxois1997}
T.~Dauxois, S.~Ruffo, A.~Torcini, {Modulational} estimate for the maximal
  {Lyapunov} exponent in {Fermi}-{Pasta}-{Ulam} chains, Phys. Rev. E 56~(6)
  (1997) R6229--R6232.

\bibitem{Cafarella2004}
A.~Cafarella, M.~Leo, R.-A. Leo, Numerical analysis of the one-mode solutions
  in the {Fermi}-{Pasta}-{Ulam} system, Phys. Rev. E. 69 (2004) 046604.

\bibitem{Antonopoulos2006IJBC}
C.~Antonopoulos, T.~Bountis, C.~Skokos, Chaotic dynamics of {$N$}-degree of
  freedom {Hamiltonian} systems, Int. J. Bif. Chaos 16~(6) (2006) 1777--1793.

\bibitem{Bountis2010}
T.~Bountis, G.~Chechin, V.~Sakhnenko, Stability of motion and discrete
  symmetries in {Hamiltonian} dynamics, Int. J. Bif. Chaos, {\rm to appear}.

\bibitem{Leo2010}
M.~Leo, R.-A. Leo, P.~Tempesta, Thermostatistics in the neighborhood of the
  $\pi$-mode solution for the {Fermi}-{Pasta}-{Ulam} $\beta$ system: {From}
  weak to strong chaos, Journal of Statistical Mechanics: Theory and Experiment
  04 (2010) 04021.

\bibitem{Cretegnyetal1998}
T.~Cretegny, T.~Dauxois, S.~Ruffo, T.~Alessandro, {Localization} and
  equipartition of energy in the $\beta$-{FPU} chain: {Chaotic} breathers,
  Physica D: Nonlinear Phenomena 121 (1998) 109--126.

\bibitem{RuizBountisTsallis2010}
G.~Ruiz-Lopez, T.~Bountis, C.~Tsallis, Time-evolving statistics of chaotic
  orbits of conservative maps in the context of the {Central} {Limit}
  {Theorem}, preprint.

\bibitem{Antonopoulos2010PRE}
C.~Antonopoulos, V.~Basios, T.~Bountis, Weak chaos and the ``melting
  transition'' in a confined microplasma system, Phys. Rev. E. 81 (2010)
  016211.

\bibitem{Hill1994}
T.-L. Hill, {Thermodynamics} of {Small} {Systems}, Dover, New York, 1994.

\bibitem{Benettin1980a}
G.~Benettin, L.~Galgani, A.~Giorgilli, J.-M. Strelcyn, Lyapunov characteristic
  exponents for smooth dynamical systems and for {Hamiltonian} systems: {A}
  method for computing all of them. {Part} 1: {Theory}, Meccanica 15 (1980)
  9--20.

\bibitem{Benettin1980b}
G.~Benettin, L.~Galgani, A.~Giorgilli, J.-M. Strelcyn, Lyapunov characteristic
  exponents for smooth dynamical systems and for {Hamiltonian} systems: {A}
  method for computing all of them. {Part} 2: {Numerical} application,
  Meccanica 15 (1980) 21--30.

\bibitem{Skokosetal2010}
C.~Skokos, E.~Gerlach, {Numerical} integration of variational equations, Phys.
  Rev. E 82~(3) (2010) 036704.
\newblock \href {http://dx.doi.org/10.1103/PhysRevE.82.036704}
  {\path{doi:10.1103/PhysRevE.82.036704}}.

\bibitem{Skokos2010}
C.~Skokos, The {Lyapunov} characteristic exponents and their computation,
  Lecture Notes in Physics 790 (2010) 63--135.

\bibitem{UmarovTsallis2008}
S.~Umarov, C.~Tsallis, S.~Steinberg, On a $q$-limit theorem consistent with
  {Nonextensive Statistical Mechanics}, Milan Journal of Mathematics 76 (2008)
  307--328.

\bibitem{Fermi1955}
E.~Fermi, J.~Pasta, S.~Ulam, Studies of nonlinear problems, Los Alamos document
  LA-1940, Addison-Wesley.

\bibitem{Fermi1974}
E.~Fermi, J.~Pasta, S.~Ulam, Nonlinear wave motion, American Mathematical
  Society, Providence, Lectures in Applied Mathematics 15.

\bibitem{Numericalrecipes}
W.~Press, S.~Teukolsky, W.~Vetterling, B.~Flanney, {Numerical} {Recipes} in
  {Fortran} 77. {The} {Art} of {Scientific} {Computing}, {Second} edition,
  Published by the Press Syndicate of the University of Cambridge.

\bibitem{Yoshida1990}
H.~Yoshida, {Construction} of higher order symplectic integrators, Phys. Let. A
  150 (1990) 262--268.

\bibitem{DeLucaetal1995}
J.~De~Luca, A.~J. Lichtenberg, S.~Ruffo, Energy transition and time scales to
  equipartition in the {Fermi-Pasta-Ulam} oscillator chain, Phys. Rev. E 51
  (1995) 2877--2885.

\bibitem{Berchiallaetal2004}
L.~Berchialla, A.~Giorgilli, S.~Paleari, Exponentially long times to
  equipartition in the thermodynamic limit, Phys. Lett. A 321 (2004) 167--172.

\bibitem{Bambusietal2008}
D.~Bambusi, A.~Ponno, Resonance, metastability and blow-up in {FPU}, Lect.
  Notes Phys. 728 (2008) 191--205.

\bibitem{Benettinetal2009}
G.~Benettin, R.~Livi, A.~Ponno, The {Fermi-Pasta-Ulam} problem: {Scaling} laws
  vs. initial conditions, J. Stat. Phys. 135 (2009) 873--893.

\bibitem{Ooyama1969}
N.~Ooyama, H.~Hirooka, N.~\Saito, Computer studies on the approach to thermal
  equilibrium in coupled anharmonic oscillators. {II}. {One} dimensional case,
  Journal of the Physical Society of Japan 27~(4) (1969) 815--824.

\bibitem{Antonopoulos2006PRE}
C.~Antonopoulos, T.~Bountis, Stability of simple periodic orbits and chaos in a
  {Fermi}-{Pasta}-{Ulam} lattice, Phys. Rev. E. 73 (2006) 056206.

\bibitem{Christodoulidi2010}
H.~Christodoulidi, C.~Efthymiopoulos, T.~Bountis, Energy localization on
  q-tori, long term stability and the interpretation of the {FPU} paradox,
  Phys. Rev. E. 81 (2010) 016210.

\bibitem{Gaspard2003}
P.~Gaspard, Lyapunov exponent of ion motion in microplasmas, Phys. Rev. E. 68
  (2003) 1--7.

\bibitem{Skokos2001}
C.~Skokos, {Alignment} {Indices}: {A} new, simple method for determining the
  ordered or chaotic nature of orbits, J. Phys. A 34 (2001) 10029--10043.

\bibitem{Skokos2003}
C.~Skokos, C.~Antonopoulos, T.-C. Bountis, M.-N. Vrahatis, How does the
  {Smaller} {Alignment} {Index} {(SALI)} distinguish order from chaos?,
  Progress of Theoretical Physics Supplement 150 (2003) 439--443.

\bibitem{Skokos2004}
C.~Skokos, C.~Antonopoulos, T.-C. Bountis, M.-N. Vrahatis, Detecting order and
  chaos in {Hamiltonian} systems by the {SALI} method, J. Phys. A 37 (2004)
  6269--6284.

\bibitem{Aubry2010}
S.~Aubry, Private communication.

\end{thebibliography}

\end{document}